\documentclass{sig-alternate} 

\usepackage{ifthen}

\usepackage[hang,flushmargin,stable]{footmisc}
\usepackage{paralist}

\usepackage{graphicx}
\usepackage{times}
\usepackage{amsfonts, amsmath}
\graphicspath{{figures/}}

\usepackage{framed}
\usepackage{flushend}

\usepackage[bf]{caption}

\usepackage[labelformat=simple]{subcaption}

\makeatletter
\let\@copyrightspace\relax
\makeatother

\usepackage{algpseudocode}
\usepackage{algorithm}

\newenvironment{argument}{\textsc{Argument}.}{\qed}

\usepackage{url}

\makeatletter
\def\url@leostyle{%
  \@ifundefined{selectfont}{\def\UrlFont{}}%
  {\def\UrlFont{}}%
}
\makeatother	
\usepackage{breakurl}

\usepackage{color,xcolor}
\definecolor{darkgreen}{RGB}{0,90,90}
\usepackage[bookmarks=false, colorlinks=true, plainpages = false,  linkcolor = darkgreen,   citecolor = darkgreen, urlcolor = darkgreen, filecolor = black]{hyperref}

\definecolor{nfvyellow}{RGB}{255,238,170}

\usepackage{enumitem}

\newtheorem{definition}{Definition}
\newtheorem{claim}{Claim}
\newcommand{\descr}[1]{\medskip\noindent\textbf{#1.}}

\begin{document}

\title{Private Processing of Outsourced Network Functions: Feasibility and Constructions\titlenote{A preliminary version of this paper appears in the 1st ACM International Workshop on
Security in Software Defined Networks \& Network Function Virtualization. This is the full version.}}

\numberofauthors{2}

\author{
\alignauthor
Luca Melis\\
       \affaddr{University College London, UK}\\
       \email{luca.melis.14@ucl.ac.uk}
\alignauthor
Hassan Jameel Asghar\\
       \affaddr{Data61, CSIRO, Australia}\\
       \email{hassan.asghar@data61.csiro.au}
\and
\alignauthor 
Emiliano De Cristofaro\\
       \affaddr{University College London, UK}\\
       \email{e.decristofaro@ucl.ac.uk}
\alignauthor 
Mohamed Ali Kaafar\\
       \affaddr{Data61, CSIRO, Australia}\\
       \email{dali.kaafar@data61.csiro.au}
}

\maketitle

\begin{abstract}
Aiming to reduce the cost and complexity of maintaining networking infrastructures, organizations are increasingly outsourcing their network functions (e.g., firewalls, traffic shapers and intrusion detection systems) to the cloud, and a number of industrial players have started to offer network function virtualization (NFV)-based solutions. Alas, outsourcing network functions in its current setting implies that sensitive network policies, such as firewall rules, are revealed to the cloud provider. In this paper, we investigate the use of cryptographic primitives for processing outsourced network functions, so that the provider does not learn any sensitive information. More specifically, we present a cryptographic treatment of privacy-preserving outsourcing of network functions, introducing security definitions as well as an abstract model of generic network functions, and then propose a few instantiations using partial homomorphic encryption and public-key encryption with keyword search. We include a proof-of-concept implementation of our constructions and show that network functions can be privately processed by an untrusted cloud provider in a few milliseconds.
\end{abstract}

\section{Introduction}
\label{sec:intro}
Network functions, such as firewalls and load balancers, are increasingly moving to ``the cloud'' by means of software processes outsourced on commodity servers. 
Using virtualization, network functions can be emulated in software in a cost-effective manner, and outsourced to the cloud reaping the benefits of reduced management and infrastructure costs, pay-per-use, etc.~\cite{aplomb}.
Specifically, network function virtualization (NFV) is currently being proposed by several major industrial operators like Cisco, Alcatel-Lucent, and Arista, as a service to multiple clients~\cite{top26}. %

In such a multi-tenancy setting, network functions are run on virtual machines (VMs) belonging to different clients hosted on the same hardware (server). Naturally, this raises a number of security concerns for clients, including confidentiality and integrity. While such issues are common to IT infrastructure outsourcing in general~\cite{cloud-visor}, more specific to NFV is the sensitivity of an organization's proprietary network policies, which instruct how network functions are to be performed. These are potentially vulnerable to compromise from competing organizations as well as the cloud service provider itself. %
For instance, firewall rules do not only reveal IP addresses of hosts, network topology, etc., but also
defense strategies and sensitivity of different services and resources, which, in the traditional setting,
are only known to a few network administrators~\cite{bf-firewall, mlm-firewall}. 
While virtual machine isolation~\cite{cloud-visor} could potentially address some of these issues, they are inadequate to provide privacy against the operator, i.e., the cloud service provider. %

\descr{Problem Statement} These challenges motivate the need to protect the privacy of network policies %
against an untrusted cloud provider, as well as other tenants and third parties. In the rest of the paper, we call this the \emph{private NFV problem}, which, as we discuss in Section~\ref{sec:related}, has been largely overlooked by prior work on NFV security. 
We define a generic model to define privacy in NFV and propose several solutions based on different cryptographic primitives such as fully homomorphic encryption, partial homomorphic encryption and public-key encryption with keyword search. The solutions result from tradeoffs between privacy and performance, and can be instantiated depending on the adversarial model, %
showing that private processing of outsource network functions is already feasible today by adapting a few existing cryptographic primitives.

\descr{Contributions} We construct an abstract model of network functions which seeks to generalize most of the network functions used in practice as well as relevant adversarial models (Sections~\ref{sec:problem} and~\ref{sec:theory}). Then, based on this abstraction, we propose three different solutions: an ideal, yet not very efficient, one based on fully homomorphic encryption, and two more practical solutions based on partial homomorphic encryption and public-key encryption with keyword search (PEKS), secure 
in two different adversarial models, which we define as {\em strong} and {\em weak} (Section ~\ref{sec:solutions}). 
Our solution against the weak adversary is also the first to include stateful network functions, e.g., a stateful firewall that keeps track of open TCP/IP connections.
Finally, we present a proof-of-concept implementation of our schemes %
and evaluate their performance overhead using an outsourced firewall as a use-case (Section~\ref{sec:implement}). Using a typical 5-tuple based firewall rule, we show that a packet can be processed within 109 ms and 180 ms, respectively, using our solutions secure against the weak and the strong adversary, and demonstrate that our schemes scale quite well, as processing times reach 250 ms and 1,208 ms, respectively, using 10 rules. Bearing in mind that our proof-of-concept implementation is not optimised for efficiency (e.g., lack of multi-threading), our results indicate that private NFV is feasible using existing cryptographic primitives.

\section{Related Work}
\label{sec:related}
Khakpour and Liu~\cite{bf-firewall} introduce a data structure called Bloom Filter Firewall Decision Diagram (BFFDD) in order to anonymize firewall policies built from Firewall Decision Diagrams (FDD)~\cite{fdd}. However, as acknowledged by the authors, Bloom filters~\cite{bloom1970space} naturally introduce false positives. Thus, occasionally, packets that do not match any policy are (mistakenly) dropped by the firewall. Furthermore, security/privacy of their solution is argued against a black-box assumption of Bloom filters, which does not analyze the security properties of Bloom filters themselves (such as one-wayness).

Shi, Zhang, and Zhong~\cite{mlm-firewall} use %
multilinear maps from Coron, Lepoint and Tibouchi (CLT), which are
based on \emph{graded encoding systems}~\cite{mlm-integers}, to encode each bit of a firewall rule as a pair of level-1 encodings and a level-$(n + 1)$ encoding for the whole rule, where $n$ is the length of a possible packet. Following the security properties of the multilinear map, it is not possible to obtain level-$i$ or lower encodings given a level-$(i+1)$ encoding for each $i$. Upon receiving a packet, the encodings corresponding to the bits of the packet are multiplied and the result is then matched with the level-$(n+1)$ encoding for the whole policy through a procedure called \texttt{isZero}. Unfortunately, the CLT construction has been recently shown to be insecure, due to an attack on the \texttt{isZero} routine~\cite{mlm-cryptanalysis}; a key ingredient to check if a packet matches a policy.

Although both these constructions focus specifically on outsourcing firewalls, they exclude details of how state tables can be maintained in their framework by a stateful firewall. Furthermore, due to being specific to firewalls, their solutions are only relevant to policies that result in a binary decision (allow or deny), excluding network functions that modify packet contents or perform more complex actions. Compared to these two solutions, our solutions for private NFV cover a much broader range of network functions, including firewalls, and also consider state tables. %

Private NFV also resembles real-time processing over encrypted packets. The work in \cite{blindbox} discusses deep packet inspection over encrypted data, however, it requires the sender (third party) to be a participant in the protocol, which makes it impossible to use this solution on existing infrastructures (a requirement that we describe as compatibility in Section~\ref{sec:desired}).

Somewhat related is work on outsourcing frameworks in Software Defined Networks. Specifically,  
Sherry et al.~\cite{aplomb} provide a prototype of the APLOMB architecture, where the middlebox functionalities (e.g. firewall) are outsourced to the cloud by the enterprises without greatly damaging throughput.
Gibb et al.\cite{waypoints} then present an architecture in which enterprise networks only forward data and additional processing is performed by external feature providers without any limitation on location. However, \cite{aplomb,waypoints} do not consider private processing.

Security issues in outsourcing network functions are also studied in, e.g.,, which provide a roadmap on the construction of a verifiable network function architecture that can verify the correctness of the outsourced service w.r.t. functionality, performance, and actual workload in the cloud.
In general, concerns raised from the the lack of control with cloud outsourcing have been investigated in~\cite{chow2009controlling}, while, addresses the problem of auditing outsourced computation by providing a monitor system that efficiently and verifiably tracks memory use and CPU-cycle consumption in the cloud.
Remote attestation and verification are also studied  by Haeberlen et al.~\cite{vu2013hybrid}, who propose an efficient method for verifying specific types of computation, while~\cite{haeberlen2010accountable} introduces accountable virtual machines without trusted hardware.
Finally, Zhang et al.~\cite{zhang2008packet} and Argyraki et al.~\cite{argyraki2010verifiable} provide mechanisms to ensure accountable networking by discovering entities that drop packets in a malicious way.

\begin{figure*}[!htb]
\centering\includegraphics[scale=0.38]{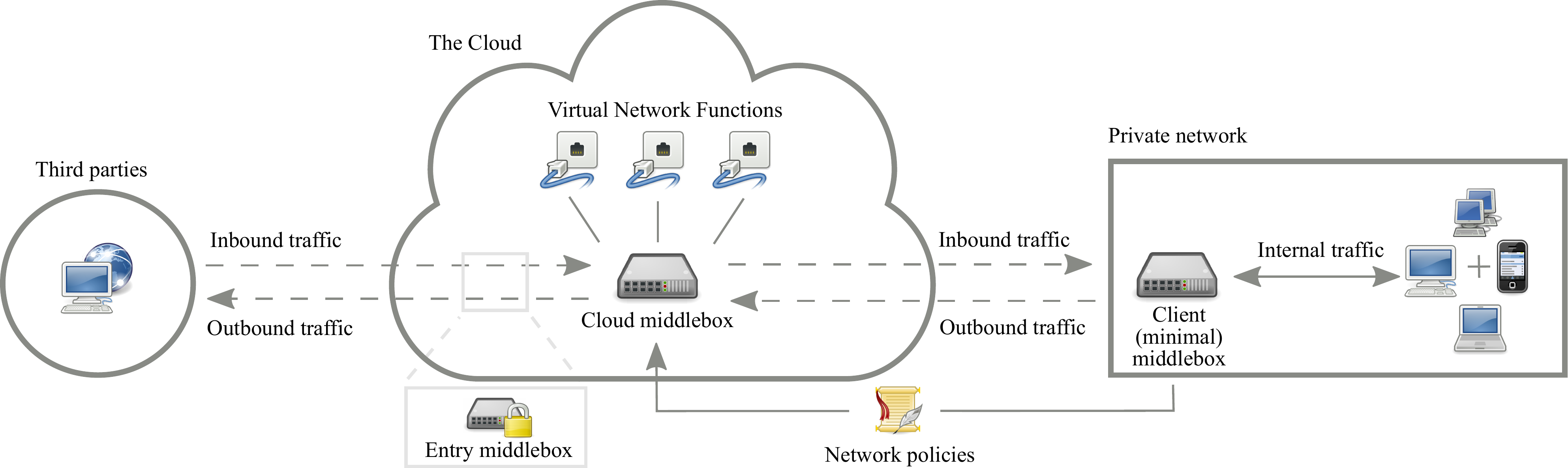}
\caption{Network Function Virtualization.}
\label{fig:nfv-diagram}
\end{figure*}

\section{Preliminaries}
\label{sec:problem}
This section introduces the problem of private processing of outsourced network functions.

\subsection{Examples of Network Functions}
In the rest of the paper, we consider outsourcing of simple network functions, 
such as those presented below, along with the related (simplified) policies.

	\descr{Firewall} The simplest example of a firewall policy is to drop a packet if the source IP address belongs to a given IP range.
	 
	\descr{Load Balancer} A load balancer distributes incoming packets across different servers to minimize load on one or more servers. A typical load distributing algorithm is round-robin. For instance, if the IP address of the server currently at the top of the list is \texttt{192.168.0.1}, then the destination IP address of the packet should be changed to this IP address. 

	\descr{Carrier-grade NAT} A carrier-grade network address translator maps private IP addresses (and ports) within a private network to one or more public IP addresses (and ports), to reduce the number of public IP addresses required. An example of a NAT policy is that if the destination IP of an incoming packet is \texttt{213.145.163.231} and the destination port is \texttt{5000}, the destination IP and port should be changed to \texttt{196.168.0.1} and \texttt{22}, respectively.

	\descr{IDS} An intrusion detection system scans packets to detect any malicious traffic. An example policy could be that if the destination IP address of an incoming packet is \texttt{192.168.0.1} and the payload contains a \texttt{POST} request then an \texttt{alert} message should be sent.

	\descr{DPI} Deep packet inspection filters packets by inspecting it for viruses or other content such as pornography. An example policy could be that if an incoming packet contains the word \texttt{adult} in its contents, then the packet should be dropped.

\subsection{System Model} 
In the rest of the paper, we consider a scenario where an organization, the \textit{client}, outsources one or more of its network functions to the \textit{cloud}, as illustrated in Figure~\ref{fig:nfv-diagram}. The outsourced network functions run within virtual machines (VMs) on commodity servers provided by the cloud. We call this the {\em NFV setting} -- as opposed to the {\em traditional} setting in which dedicated network middleboxes perform network functions within the client's private network.  Analogous to other cloud platforms, such VMs are managed through hypervisors~\cite{click-os}. 

\descr{Cloud and Client Middleboxes} To ease presentation, we denote the set of all VMs executing virtual network functions as the {\em cloud middlebox}, or {\em cloud MB} for short. Not all network functionalities need to be outsourced to the cloud, and as such the client still requires its own middlebox to carry out the remaining network functions or to communicate with the cloud MB. We call this the {\em client middlebox}, or {\em client MB} for short. The cloud MB receives inbound traffic destined for the client, processes the network functions assigned to it, and forwards the result to the client MB. Outbound traffic is the one originating from within the client's private network which is forwarded by the client MB to the cloud MB to process the outsourced network functions and subsequently relay it to its intended destination. The network policies which describe how the network functions are to be processed are installed in the cloud MB by the client. 

\descr{Trust Assumptions} We assume the cloud MB to be honest-but-curious, i.e., it performs network functions dutifully yet wishes to infer the policies. Later on in this paper, for some of the proposed solutions, we will assume that the cloud MB has a \emph{semi-trusted} component, which we call the \emph{entry MB}. The entry MB receives the packet and performs some preliminary processing before handing the results over to the cloud MB. Ideally there should be no entry MB, i.e., no part of the cloud MB should be assumed to be part of black-box processing. However, inclusion of an entry MB remarkably improves performance, and its presence is reasonable assuming that the cloud is honest-but-curious. Also, remark that the entry MB does not share any private keys with the client MB, and all the processing is done using public-key operations. 
\subsection{Desired Properties}\label{sec:desired}
In the traditional setting, most network functions are run on dedicated middleboxes located at the edge of the client's private network. As a result, the network policies are hidden from outsiders as long as the hardware is secure. Once a network function is outsourced to the cloud, obviously, it is no longer the case. Ideally, the client would want its network policies to remain private while maintaining the standards of service set by the traditional setting. %

\descr{Privacy} The client expects its network policies to remain hidden not only from 
third parties, but also from other tenants and the cloud. %
We argue that the cloud should not be trusted to keep the policies secret, even though it processes the network functions for the client. At best, the client can only assume that the cloud is {\em honest-but-curious}, i.e., it performs all the network functions as required due to service obligations and does not deviate from protocol specification, but it might still be interested in inferring network policies, possibly by colluding with another party. %
Also, due to virtualization, it is likely that two VMs computing network functions of two (possibly competing) tenants might be residing in the same physical server, thus, a client's network policies should be kept secret from another client.

\descr{Performance} The client expects the outsourced network functions to maintain the quality of service of the traditional setting. This introduces the following constraints.
\begin{itemize}
	\item \textit{Real-time Processing}: The cloud MB should be able to process network functions in real-time.
	\item \textit{Minimal Client-side Processing}: The client MB should be processing as little of the policies as possible in order to maintain the benefits of network function outsourcing. %
\end{itemize}

\descr{Compatibility} Third parties should be able to send/receive traffic to/from the client as if the network functions are implemented in the traditional setting, i.e., third parties should not be required to undergo additional setup (e.g., implementation of customized network and cryptographic protocols) %
to communicate with the client.
\smallskip

Naturally, any solution for a private NFV will likely introduce a tradeoff between  privacy and compatibility/performance: our goal is to explore the balance between security and performance, while satisfying the compatibility constraint.

\subsection{Limitations and Scope}
Before introducing our solutions, we discuss a few limitations of our model and make some important remarks.

\descr{Traffic Analysis} An adversary may intercept and analyze traffic between the cloud MB and a third party and try to infer network policies based on the pattern of inbound and outbound packets. Likewise, the adversary may generate its own traffic destined for the client (through the cloud MB) and analyze the packets it receives in response. For instance, if a request has been sent from a certain IP address for a TCP/IP connection, and a response has not been received, then the adversary may infer that it is a policy to drop packets from this particular IP address. However, note this can also be done in the traditional setting, and we require that solutions for private NFV do not need to provide privacy beyond what can be achieved in the traditional setting. 

\descr{Virtual Machine Isolation} One way to achieve private NFV is through VM isolation, e.g., isolation of memory and disk storage, together with the assumption that the hypervisor belongs to a trusted base~\cite{nohype, sel4, cloud-visor}. A crucial aspect for secure isolation is to ensure that the hypervisor, i.e., trusted computing base, is small in terms of lines of codes (LoC)~\cite{hypersafe,cloud-visor}, which ensures that security vulnerabilities are minimized or, if identified, can be easily patched~\cite{sel4}. There are, however, several issues with this approach. %
(1)~Small hypervisors are needed to formally verify correctness and security properties, and some simplifying assumptions are required, e.g., w.r.t. the correctness of the compiler and the hardware, the presence of a uniprocessor instead of multiprocessors, etc., as in the case of the formal verification of the operating system kernel ``sel4''~\cite{sel4}. Also, it may be possible to iteratively verify a hypervisor by shedding each layer of simplifying assumptions. (2)~Unfortunately, commodity hypervisors are not optimized in terms of lines of codes~\cite{cloud-visor}, thus it is a strong assumption to assume they are trusted. (3)~Cross-VM side-channel attacks can also enable a malicious VM to be co-located at the physical host of the target VM and exploit various side channels (e.g., cache), to obtain information such as cryptographic keys~\cite{hey-you, zhang-juels}. 

\descr{Coverage of Network Functions} Our goal is to provide solutions to private NFV that are applicable to most network functions, ideally, encompassing all possible network functions. However, one cannot make such claim
without checking the implementation details of each and every network function in practice. Rather, we give a broad definition of network functions and provide solutions to private NFV that cover network functions satisfying this definition, which can be incrementally modified to cover more functions. For instance, we do not consider traffic shaping, where delivery of certain packets is delayed (at the cloud's side) to satisfy performance guarantees. 

\descr{Inbound vs Outbound Traffic} In this work, we focus on {\em inbound} traffic, i.e., traffic coming from third parties toward the client. Although our private NFV solutions (presented next) are applicable to outbound traffic as well, this would require redirecting traffic from the cloud MB (after private processing of network functions) to the client MB, which in turn forwards it to the third party receiver. %
\section{Mathematical Formulation}
\label{sec:theory}

Let $n$ be a positive integer and $\mathbf{x}$ and $\mathbf{y}$ be $n$-element vectors: then
$\langle \mathbf{x}, \mathbf{y} \rangle$ denotes their dot product. The dot product of a vector $\mathbf{x}$ with itself, i.e., $\langle \mathbf{x}, \mathbf{x} \rangle$ is denoted by $\mathbf{x}^2$. The Hadamard product or the entry-wise product of the vectors $\mathbf{x}$ and $\mathbf{y}$ is $\mathbf{x} \circ \mathbf{y}$,
i.e., the $n$-element vector whose $i$-th element is $x_iy_i$. The vector $\mathbf{e}_i$ denotes the $n$-element vector with all $0$s except a $1$ in the $i$-th position.

Given two positive integers $a$ and $b$, the \emph{bitwise AND} operation, denoted $a \odot b$, outputs $1$ if the binary representation of $a$ and $b$ agrees in all bit positions. More specifically, if we assume $a$ and $b$ to be $n$-bit binary numbers and let $a_i$ and $b_i$ denote their $i$-th bits with the most significant bit at position $n$, then
\[
a \odot b = c_n c_{n-1} \cdots c_1, 
\]
where $c_i = a_ib_i +  \bar{a}_i \bar{b}_i$. The \emph{bitwise greater than or equal to} operation, denoted $a \unrhd b$, is defined as
\begin{align*}
a \unrhd b &= a_n\bar{b}_n + c_n a_{n-1}\bar{b}_{n-1} + c_n c_{n-1} a_{n-2}\bar{b}_{n-2} + \cdots \\
				  &+ c_n \cdots c_2 a_{1}\bar{b}_{1} + c_n c_{n-1} \cdots c_1,
\end{align*}
which is $1$ if $a \ge b$ and $0$ otherwise. The \emph{bitwise less than or equal to} operation, denoted $a \unlhd b$, is defined similarly with the roles of $a$ and $b$ interchanged. 

The encryption function $E$ on a vector $\mathbf{x}$ is defined as the vector
\[
E(\mathbf{x}) = 
\begin{pmatrix}
E(x_1) & E(x_2) & \cdots & E(x_n)
\end{pmatrix}.
\]

For positive integers $a < b$, the notation $[a, b]$ denotes all integers between $a$ and $b$ inclusive. The notation $[n]$, for a positive integer $n$, defines the set $\{1, 2, \ldots, n\}$.

\subsection{Network Functions}
Let $n \ge 1$ and $q \ge 2$ be positive integers. We define a packet $\mathbf{x}$ as a vector in $\mathbb{Z}^{n}_q$, where $n$ represents the different fields of the packet (source IP address, protocol type, etc.) and $q$ is an upper bound on the length of packet fields. Although it is much natural to define a packet as a bit string of bounded length ($2^{16}$ in case of IPv4 packets), we prefer our definition as it facilitates the description of private NFV solutions later on. A network function $\psi$ from $\mathbb{Z}^{n}_q$ onto $\mathbb{Z}^{n}_q$ is the pair $(m, a)$ defined as
\begin{equation}
\label{eq:nf}
\psi(\mathbf{x}) = m(\mathbf{x})a(\mathbf{x}) + (1 - m(\mathbf{x}))\mathbf{x}, 
\end{equation}
where $m : \mathbb{Z}^{n}_q \rightarrow \{0, 1\}$ is called the matching function, and $a: \mathbb{Z}^{n}_q \rightarrow \mathbb{Z}^{n}_q$ is the action function, or simply the {\em action}. The intuitive meaning of the above is that when a network function receives a packet $\mathbf{x}$ the matching function decides whether the current network function applies to this packet. If yes, the relevant action is performed by the action function altering the packet to $\mathbf{x}'$. If the result of the match is negative, the packet is left unchanged. 

In some cases, a network policy will be composed of several network functions as defined above -- in this case, we iteratively define the resulting network function as:
\begin{equation}
\label{eq:nf-comp}
\psi^{i}(\mathbf{x}) = \psi_i(\cdots\psi_2(\psi_1(\mathbf{x}))\cdots)
\end{equation}
for $i \ge 1$. 

The definition of $\psi$ as a match-action pair is motivated by the OpenFlow communications protocol between the control and forwarding planes in Software Defined Networks (SDN)~\cite{openflow}, which use flow tables containing match fields and the corresponding actions to be carried out. 
Note that different fields of a packet are not necessarily of the same length, e.g., if we consider IP packets then the version field (i.e., IPv4 or IPv6) is 4 bits long while the source IP field is 32 or 128 bits long (IPv4 or IPv6 packets). Therefore, we consider a value $q$ that is large enough to incorporate the largest header field.
This is for theoretical convenience, and any superfluous bits for smaller fields can be duly discarded. The packet payload, which can be much larger, is divided into chunks of length $\log_2 q$ bits. 

\descr{Virtual Fields} Besides the standard fields, we assume the presence of additional ones, which we call \emph{virtual} fields. These  originate from the implementation of our private NFV instantiations and are inserted in the payload of the packet. For instance, a \emph{tag} field will be used to model a common functionality of network functions such as the firewall and rate limiter, to drop packets matching certain criteria. To indicate that a packet is to be dropped, the cloud MB can assign the value \texttt{drop} to this tag (contained in the IP packet's payload) and send it to the client MB. How this value is added in a private way is described in Section~\ref{sec:solutions} and how these virtual fields can be added to the packet is described in Section~\ref{sec:implement}.

\descr{Example}
We assume a simple network address translation (NAT) policy as a running example. For instance, upon receiving a packet $\mathbf{x}$ with destination IP in the range \texttt{128.*.*.*}, the NAT changes the destination IP and port to \texttt{196.*.*.*} and \texttt{22}, respectively. Without loss of generality, we assume that the destination IP and destination port belong to the first two elements of $\mathbf{x}$, i.e., $x_1$ and $x_2$. Thus, the matching function is: 
\[
m(\mathbf{x}) = \begin{cases}
				1 & \text{if } x_1 \in [\texttt{128.0.0.0}, \texttt{128.255.255.255}]\\
				0 & \text{otherwise}
				\end{cases},
\]
and the action is:
\[
a(\mathbf{x}) = 
\begin{pmatrix}
x_1 \\
x_2 \\
x_3 \\
\vdots \\
x_n			
\end{pmatrix} +
\begin{pmatrix}
x_1 + \texttt{68.0.0.0} \\
-x_2 + 22 \\
0 \\
\vdots \\
0			
\end{pmatrix}.
\]
(Note that the IP addresses are mapped in $\mathbb{Z}_q$.)

\subsection{Stateful Network Functions}
Some network functions such as (stateful) firewalls maintain dynamically generated states. When a packet arrives, it is first checked against the state table to see if any entry in the state table matches the fields of the packet. If a matching entry is found, the prescribed action is performed on the packet and it does not need to be further processed by other (static) policies. An example is the state of TCP connection maintained by a firewall, as depicted in Table~\ref{table:state}. The firewall notes a \textit{new} connection when the SYN flag in a packet is set, and creates an entry in the state table containing, for instance, the source/destination IPs and ports and protocol type (6 for TCP) along with the current state. Upon receiving a SYN-ACK from the destination and a subsequent ACK flag from the source, the firewall changes the state of this connection to \textit{established} (shown as \texttt{est} in the table). Any subsequent packets that satisfy the headers contained in the state table entry are then allowed to go through without further processing of the network policies. The state table entry is deleted once the FIN-ACK part of the TCP protocol is carried out, or when the connection times out. 

\begin{table}[!h]
\centering
\resizebox{1.025\linewidth}{!}{
\setlength\tabcolsep{2pt} %
\begin{tabular}{l|l|r|l|r|r|r|r} 
ID   & src IP & src port  & dst IP & dst port  & prot  & state & timeout  \\
\hline
1 & \texttt{192.168.1.1} & \texttt{120} & \texttt{192.168.1.2} & \texttt{121} & \texttt{6} & \texttt{new} & \texttt{59}\\
2 & \texttt{192.168.1.129}  & \texttt{45} & \texttt{192.168.1.140}   & \texttt{8080} & \texttt{6} & \texttt{est} & \texttt{3600}
\end{tabular}
}
\vspace{-0.1cm}
\caption{An example of a firewall state table.}
\label{table:state}
\vspace{-0.1cm}
\end{table}

We note that in our model, state tables can be abstracted as dynamic match-action pairs, where the state and time-out columns in the state table can be thought of as \textit{virtual} fields of the IP packet and the action as the addition of the \texttt{tag} field with value ``allow''. However, one key difference is that once a match has been found, further processing is discontinued.\footnote{There are network functions for which this is not true, e.g., traffic monitoring in which aggregate statistics of packets, such as number of packets received, are maintained.} Therefore, any private solution to a stateful middlebox should have the property that execution is allowed to stop once a match in the state table is found---otherwise there would be no performance gain from maintaining state.

\subsection{Private Processing of Outsourced Network Functions}
Our goal is to provide privacy of an outsourced network function $\psi$ given a set of packets $\mathbf{x}_1, \mathbf{x}_2, \ldots, \mathbf{x}_t$. From an adversarial perspective, the network function $\psi$ can be learned either directly through the description of $(m, a)$ or indirectly by deducing from the outputs $\psi(\mathbf{x}_1), \psi(\mathbf{x}_2), \ldots, \psi(\mathbf{x}_t)$. In order to achieve privacy, we therefore need a scheme that protects both the network function $\psi$ and its output. We call this PNFV (Private NFV). Let $\mathbf{x}$ be a packet as defined before and $\psi$ be a network function such that $\psi(\mathbf{x}) = \mathbf{x}'$.

\subsubsection{PNFV}
\vspace{-0.25cm}
\begin{definition}[PNFV]
A public-key PNFV scheme is a tuple 
$(\mathsf{kg}, \mathsf{enc}, \mathsf{dec}, \mathsf{tr}, \mathsf{proc})$
of probabilistic polynomial time algorithms defined as follows:
\begin{compactitem}
\item {\em Key generation:} The algorithm $s, p \leftarrow \mathsf{kg}(1^k)$ returns the secret key $s$ and public key $p$, where $k$ is the security parameter.
\item {\em Packet encryption:} The algorithm $E(\mathbf{x}) \leftarrow \mathsf{enc}(p, \mathbf{x})$ takes as input the public key $p$ and the packet $\mathbf{x}$ and outputs the encrypted version $E(\mathbf{x})$. Note that this is element-wise encryption, which results in $n$ ciphertexts.
\item {\em Network function transformation:} The algorithm $\phi \leftarrow \mathsf{tr}(\psi)$ takes as input the network function $\psi$ and outputs a \emph{transformed} network function $\phi$. 
\item {\em Packet processing:} The algorithm $E(\mathbf{x}') \leftarrow \mathsf{proc}(\phi, E(\mathbf{x}))$ takes as input the transformed network function $\phi$ and the encrypted packet $E(\mathbf{x})$ and outputs the encryption of $\mathbf{x}'$. 
\item {\em Packet decryption:} The algorithm $\mathbf{x}' \leftarrow \mathsf{dec}(s, E(\mathbf{x}'))$ takes as input the secret key $s$ and the encryption of $\mathbf{x}'$ and outputs $\mathbf{x}'$. We may write $D(E(\mathbf{x}))$ to represent $\mathsf{dec}(s, E(\mathbf{x}))$.
\end{compactitem}
\end{definition}
Concisely, we can define the output of PNFV given $\mathbf{x}$ and $\psi$ as $\text{PNFV}(\mathbf{x}, \psi)$. Thus,
\[
\text{PNFV}(\mathbf{x}, \psi) = \mathsf{dec}(s, \mathsf{proc}(\mathsf{tr}(\psi), \mathsf{enc}(p, \mathbf{x}))).
\]
Key generation, network function transformation, and packet decryption algorithms are computed by the client MB, while the remaining two algorithms are processed by the cloud MB. We have the following definition for correctness.

\begin{definition}[Correctness]
A \emph{public-key} {\em PNFV} scheme is correct if for all $\mathbf{x} \in \mathbb{Z}_q^n$ it holds that  
\[
\Pr [ \text{\emph{PNFV}}(\mathbf{x}, \psi) \ne \psi(\mathbf{x}) ] \le \mathsf{negl}(k),
\]
where $s, p \leftarrow \mathsf{kg}(1^k)$, $\mathsf{negl}$ is a negligible function and $k$ is the security parameter.
\end{definition}
\subsubsection{PNFV Security}
As mentioned before, we consider an honest-but-curious adversary, i.e., a passive adversary that correctly computes PNFV but would like to infer $\psi$. More precisely, we conduct the following experiment involving an adversary $\cal{A}$ to model PNFV security.
First, $\cal{A}$ is given the public key $p$, the description of algorithms $(\mathsf{kg}, \mathsf{enc}, \mathsf{dec}, \mathsf{tr}, \mathsf{proc})$ and the transformed network function $\phi$. 
While $\cal{A}$ is in the \emph{test} state, it can sample any packet $\mathbf{x}$ and obtain its output $E(\mathbf{x}')$ such that $\psi(\mathbf{x}) = \mathbf{x}'$ through the packet processing algorithm. 
Finally, in the \emph{guess} state $\cal{A}$ outputs its guess of the network function $\psi$ as $\psi'$. If $\psi' = \psi$, $\cal A$ wins.
The above experiment abstracts what we call the strong adversary, denoted $\cal{A}_\text{strong}$, to distinguish it from a weaker adversary, denoted $\cal{A}_\text{weak}$. 
The weak adversary differs from the strong one in that it is only given \emph{oracle (black box) access to part} of the packet processing algorithm $\mathsf{proc}$, and is not shown the incoming packet $\mathbf{x}$. Instead a packet is chosen randomly from a publicly known distribution $\cal D$, whenever $\cal{A}_\text{weak}$ requests for outputs of the above functions on a fresh input $\mathbf{x}$. Naturally, this yields a weaker security definition.
In practice, this model is realized by introducing an entry MB, which is assumed to be running within a black box. The entry MB receives the packet and performs part of the packet processing algorithm $\mathsf{proc}$, which is hidden from $\cal{A}_\text{weak}$. 

We model PNFV security using the following experiment involving,
as discussed in Section~\ref{sec:desired}, an {\em honest-but-curious} adversary.

\begin{definition}
A \emph{public-key PNFV} scheme is $(\tau, \epsilon)$-private if for any adversary $\mathcal{A}$ that runs in time $\tau = \mathsf{poly}(k)$, it holds that
\[
\Pr [ \mathcal{A}^{\text{\emph{PNFV}}} =\psi ] \le \epsilon = \epsilon(k), 
\]
where $\mathbf{x}' = \psi(\mathbf{x})$, $\cal{A}$ can be either $\cal{A}_\text{\emph{strong}}$ or $\cal{A}_\text{\emph{weak}}$ and $k$ is the security parameter.
\end{definition}

\subsection{Notation}

In Table~\ref{tab:notation}, we summarize the notation used throughout the rest of the paper.

\begin{table}[t]
\centering
\small
\begin{tabular}{|l|l|}
\hline 
{\bf Symbol} & {\bf Description} \\ \hline \hline
$n$		& Number of packet fields		\\ \hline
$N$		& Number of policies		         \\ \hline
$x_i$ 		& $i$-th packet field \\ \hline
$\psi()$	& Network function				 \\ \hline
$m()$	& Matching function				 \\ \hline
$a()$	& Action function				 \\ \hline
$\langle \mathbf{x} , \mathbf{y}\rangle$ & Dot product of vectors $x$ and $y$ \\ \hline
$\mathbf{x} \circ \mathbf{y}$ & Entry-wise product of vectors $x$ and $y$ \\ \hline
$x \odot y$ & \emph{Bitwise AND} operation \\ \hline
$\mathbf{e}_i$ & Vector with all $0$'s and a $1$ at the $i$-th position \\ \hline
$E()$ & Encryption function \\ \hline
$D()$ & Decryption function \\ \hline
$\sigma()$ & Pseudorandom Permutation \\ \hline
$I$ & Index vector whose $i$-th element is $i$ itself \\ \hline
$||$ & Concatenation operator \\ \hline 
$\mathcal{E}()$ & Searchable encryption function \\ \hline 
$T()$ & Trapdoor generation function \\ \hline 
$\mathsf{test}()$ & Test equality function \\ \hline
\{\texttt{new},\texttt{est}\} & Set of states \\ \hline
\{\texttt{allow},\texttt{drop}\} & Set of tags \\ \hline
\texttt{id} & Identifier of a table entry \\ \hline 
\texttt{delete} & Command to delete a table entry \\ \hline 
\end{tabular}
\vspace{-0.2cm}
\caption{Notation}
\label{tab:notation}
\vspace{-0.2cm}
\end{table}

\section{Three PNFV Instantiations}
\label{sec:solutions}

This section presents three PNFV instantiations. First, we briefly review a few different cryptographic primitives used in our schemes. Then, we describe solutions for a generic network function $\psi$, which, given a packet $\mathbf{x}$ implements the policy:
\begin{align}
\tag{\texttt{P1}}
\label{eq:equal-pol}
\texttt{if } x_i == y \texttt{ then } x_j \leftarrow z,
\end{align}
where $i, j \in [n]$. We call this the \emph{equality matching} policy, a special case of the more general \emph{range matching} policy defined as:
\begin{align}
\tag{\texttt{P2}}
\label{eq:range-pol}
\texttt{if } x_i \in [a, b] \texttt{ then } x_j \leftarrow z.
\end{align}
Note that policy~\ref{eq:range-pol} equals policy~\ref{eq:equal-pol} if $a = b = y$.

\subsection{Cryptographic Primitives}

\descr{Fully Homomorphic Encryption (FHE)}
A FHE scheme involves the following algorithms:
\begin{itemize}
\itemsep0em 
\item {\em Key generation:} Given the security parameter $k$, generates public and private key pair $(pk,sk)$.
\item {\em Encryption:} Given plaintext $m \in \{ 0 , 1 \}^*$, outputs ciphertext $c = E(m)$ encrypted under public key $pk$.
\item {\em Decryption:} Given a ciphertext $c$, outputs the plaintext $m=D(c)$ using the secret key $sk$.
\item {\em Homomorphic Addition (Add):} Given two ciphertexts $c_1 = E(m_1)$, $c_2 = E(m_2)$, and the public key $pk$,  produces a ciphertext $c = \textrm{Add}(c_1,c_2)=c_1 + c_2$ such that $D(c) = m_1 + m_2$.
\item {\em Homomorphic Multiplication (Mult):} Given two ciphertexts $c_1 = E(m_1)$, $c_2 = E(m_2)$, and the public key $pk$, produces a ciphertext $c$ as $c = \textrm{Mult}(c_1,c_2)=c_1 \cdot c_2$ such that $D(c) = m_1 \cdot m_2$.
\end{itemize}

\descr{The BGN Cryptosystem~\cite{bgn}}
Boneh, Goh, and Nissim (BGN) cryptosystem~\cite{bgn}, besides provide additive homomorphism, also allows for \emph{one} multiplication of ciphertexts. The scheme is based on a bilinear map and involves the following algorithms: 
\begin{itemize}
\itemsep0em
	\item {\em Key Generation:} Generate the tuple $(q_1, q_2, {G}_1, {G}_2, e)$, where ${G}_1$ and ${G}_2$ are two multiplicative cyclic groups of order $n = q_1q_2$ and $e$ is the bilinear map $e: {G}_1 \times {G}_1 \rightarrow {G}_2$. Further pick two random generators $g$ and $u$ of ${G}_1$ and set $h = u^{q_2}$. It follows that $h$ is a random generator of the subgroup of ${G}_1$ of order $q_1$. The public key is $p = (n, G_1, {G}_2, e, g, h)$ and the private key is $s = q_1$. 
	\item {\em Encryption:} Assume the message space to be $\{0, 1, 2, \ldots, M\}$ where $M < q_2$. Encryption of a message $m$ using public key $p$ is $c = g^m h^r$, where $r$ is randomly chosen from the set $\{0, 1, \ldots, n-1\}$. $c$ is the resulting ciphertext and is an element of ${G}_1$.
	\item {\em Decryption:} Given the secret key $s = q_1$, compute $c^{q_1}$ and then find its discrete log base $g^{q_1}$ using, for instance, Pollard's lambda method which takes expected time $O(\sqrt{M})$~\cite[\S 3, p. 128]{handbook-ac}\cite{bgn}.
\end{itemize}
The BGN cryptosystem is semantically secure under the \emph{subgroup decision} assumption,
i.e., given an element $x \in G_1$, it is hard to decide if $x$ is in a subgroup of $G_1$ without knowing the factorization of the group order $n$ (which is $q_1q_2$).
Since decryption involves computing discrete logarithms, BGN is only suitable for a small message space.

\descr{PEKS~\cite{peks}}
Public-key Encryption with Keyword Search (PEKS)~\cite{peks} involves the following algorithms:
\begin{itemize}
\itemsep0em
	\item {\em Key generation:} Given a security parameter, generates the public key $p$ and private key $s$.
	\item {\em PEKS generation:} Given a keyword $w$ and the public key $p$, produces the searchable encryption $\mathcal{E}$ of $w$ as $\mathcal{E}(w)$.
	\item {\em Trapdoor generation:} Given the private key $s$ and a keyword $w$, generates the trapdoor for $w$ as $T(w)$.
	\item {\em Test:} Given public key $p$, searchable encryption $\mathcal{E}(w)$ and trapdoor $T(w')$, $\mathsf{test}(\mathcal{E}(w), T(w'))$ outputs $1$ if $w' = w$ and $0$ otherwise.
\end{itemize}
We consider the instantiation by Boneh et al.~\cite{peks}, based on Identity Based Encryption (IBE)~\cite{ibe}, which itself is based on a bilinear map $e: {G}_1 \times {G}_1 \rightarrow {G}_2$, where both $G_1$ and $G_2$ are of prime order $p$. The resulting scheme is semantically secure against a chosen-keyword attack in the random oracle model under the Bilinear Diffie-Hellman (BDH) assumption~\cite{peks, ibe}, i.e., that given $g$, $g^a$, $g^b$,  $g^c$ $\in G_1$, where $g$ is a generator of $G_1$, it is hard to compute $e(g, g)^{abc} \in G_2$. Apart from this assumption, we also use the assumption that the trapdoor $T(w)$ is not invertible, i.e., is one-way. In the specific construction discussed, the trapdoor $T$ is computed as $T(w) = H(w)^s$, where $H$ is a hash function. The one-wayness of the trapdoor follows from the one-wayness of $H$. 

\descr{Pseudorandom Permutation}
We also assume the existence of a secure pseudorandom permutation $\sigma$, mapping from $[n]$ to itself. In practice, this can be implemented using a block cipher~\cite{luby-rackoff}, such as AES. In our constructions, the inverse permutation $\sigma^{-1}$ is not required and as such the private key does not need to be shared. 
\subsection{Privacy against the Strong Adversary}
\label{sec:strongAdv}

\begin{figure*}[!htb]
\begin{framed}
\small
\textbf{\em Key generation.} The client MB creates a public-private key pair $(pk, sk)$ of the FHE scheme. It keeps $sk$, and sends the public key $pk$ to the cloud MB.

\medskip
\textbf{\em Network function transformation.} The client MB computes the encrypted tuple $(E(\mathbf{e}_i), E(\mathbf{e}_j), E(y), E(z))$ using the public key $pk$ and sends them to the cloud MB.  

\medskip
\textbf{\em Packet encryption.} Upon receiving a packet $\mathbf{x}$, the cloud MB computes $E(\mathbf{x})$, using the public key $pk$.

\medskip
\textbf{\em Packet processing.} Using $E(\mathbf{x})$ and the encrypted tuple, the cloud MB computes $E(\mathbf{x}') = E(\psi(\mathbf{x}))$ as defined in Eq.~\ref{eq:fully-nf}. 

\medskip
\textbf{\em Packet decryption.} Upon receiving the encrypted packet $E(\mathbf{x}')$, the client MB decrypts it using its private key $sk$ to obtain the transformed packet $\mathbf{x}'$.
\end{framed}
\vspace{-0.4cm}
\caption{PNFV scheme based on Fully Homomorphic Encryption (FHE). $E()$ denotes the encryption function of an FHE cryptosystem.}
\label{fig:fhesol}

\end{figure*}

\subsubsection{Scheme based on Fully Homomorphic Encryption}

We now introduce our fist solution based on FHE that is secure against the strong adversary.
Consider a  network function $\psi$ implementing policy~\ref{eq:equal-pol}. Its matching function can be written as:
\[
m(\mathbf{x}) = \langle \mathbf{x} , \mathbf{e}_i \rangle \odot y
\]
which returns $1$ if $y = x_i$ and $0$ otherwise. The action function can be written as:
\[
a(\mathbf{x}) = \mathbf{x} - \mathbf{x} \circ \mathbf{e}_j + z \mathbf{e}_ j,
\]
replacing $x_j$ with $z$. %
Thus, $\psi$ becomes:
\begin{align}
\label{eq:fully-nf}
\psi(\mathbf{x}) &= m(\mathbf{x})a(\mathbf{x}) + (1 - m(\mathbf{x}))\mathbf{x} \nonumber\\
					 &= m(\mathbf{x})(a(\mathbf{x}) - \mathbf{x}) + \mathbf{x} \nonumber\\
					 &= (\langle \mathbf{x} , \mathbf{e}_i \rangle \odot y)(z \mathbf{e}_ j - \mathbf{x} \circ \mathbf{e}_j) + \mathbf{x}.
\end{align}

We can construct a public-key PNFV scheme from any Fully Homomorphic Encryption (FHE) scheme, as described in Figure~\ref{fig:fhesol}. %

For policy~\ref{eq:range-pol}, the action function is the same, but the matching function is now given as:
\[
m(\mathbf{x}) = (\langle \mathbf{x} , \mathbf{e}_i \rangle \unrhd a )(  \langle \mathbf{x} , \mathbf{e}_i \rangle \unlhd b), 
\]
which is $1$ if $x_i \in [a, b]$ and $0$ otherwise. This can be substituted for $m(\mathbf{x})$ in Eq.~\ref{eq:fully-nf} to get an expression for $\psi$. The client MB needs to send the encryptions $E(a)$ and $E(b)$ (instead of $E(y)$) to the cloud MB, while the rest is the same. Since policy~\ref{eq:equal-pol} equals policy~\ref{eq:range-pol} with $a = b = y$, we can replace the matching function of the former with the latter for a more general description, even though incurring more homomorphic computations. Also note that one can sequentially process $N$ network functions $\psi_1 , \ldots, \psi_N$ using this scheme, with the client MB sending encryptions for each network function at setup, and the cloud MB sending the encryption of
\[
\psi^{N}(\mathbf{x}) =  \psi_N(\cdots \psi_2(\psi_1(\mathbf{x}))\cdots),
\]
to the client MB upon receiving the packet $\mathbf{x}$. 

\descr{Correctness} It is straightforward to see that, if the underlying FHE scheme is correct, the construction in Figure~\ref{fig:fhesol} correctly performs the network function defined by policies~\ref{eq:equal-pol} and \ref{eq:range-pol}. 

\descr{Privacy} Intuitively, privacy of the scheme stems from the fact that, as matching and action functions, together with their results, are encrypted, the adversary cannot infer the network function. More formally, in
Appendix~\ref{app:sec-red},
we prove that this scheme is private against $\cal{A}_\text{strong}$ if the FHE scheme is semantically secure.

\descr{FHE Practicality} Although research in FHE has made tremendous progress in improving efficiency~\cite{DBLP:conf/ccs/NaehrigLV11}, we do not have a truly efficient FHE instantiation providing acceptable performance in the context of network function virtualization. However, efficient partial homomorphic encryption schemes, like BGN~\cite{bgn}, could be used, as discussed next, if we modify the matching function.

\subsubsection{Scheme based on BGN Cryptosystem}
\label{sub:bgn-scheme}
As for the FHE based scheme, we start with the function $\psi$ described by policy~\ref{eq:equal-pol},
but describe the matching function as:
\[
m(\mathbf{x}) = 1 - \langle \mathbf{x},\mathbf{e}_i\rangle + y.
\]
If we denote $m(\mathbf{x}) = c$, note that $c = 1$ if $y = x_i$, whereas, if $x_i \ne y$, then $c \ne 1$. 
Since we only get $c$ as a function of $\mathbf{x}$, %
the matching function will output $1$ only if the packet matches the policy and give %
any value other than $1$ otherwise. 

The action function $a$ is the same as before:
\[
a(\mathbf{x}) = \mathbf{x} - \mathbf{x} \circ \mathbf{e}_j + z\mathbf{e}_j.
\]

\descr{Matching and Action} We need an encryption algorithm $E$ that can homomorphically compute both $m$ and $a$. More specifically, $E$ should give the encryption of $m()$ as:
\begin{align}
\label{eq:bgn-match}
E(m(\mathbf{x})) &= E(1 - \langle\mathbf{x},\mathbf{e}_i\rangle + y) \nonumber\\
						& = E(1) - E(\langle \mathbf{x}, \mathbf{e}_i \rangle) + E(y) \nonumber \\
						& = E(1) - \langle E(\mathbf{x}), E(\mathbf{e}_i) \rangle + E(y) 
\end{align}
and, for the action function: %
\begin{align}
\label{eq:bgn-action}
E(a(\mathbf{x})) &= E(\mathbf{x} - \mathbf{x} \circ \mathbf{e}_j + z\mathbf{e}_j) \nonumber\\
						&= E(\mathbf{x}) - E(\mathbf{x} \circ \mathbf{e}_j) + E(z\mathbf{e}_j) \nonumber\\
						& = E(\mathbf{x}) - E(\mathbf{x}) \circ E(\mathbf{e}_j) + E(z\mathbf{e}_j).
\end{align}

The BGN cryptosystem allows to homomorphically compute one multiplication and any number of additions. Therefore, we can use it to construct a PNFV scheme secure against the strong adversary: the scheme is presented in Figure~\ref{fig:bgnsol}. %
We omit the description of the key generation algorithm (which should be obvious from the underlying cryptosystem), and further include the packet encryption routine within the packet processing algorithm. 

\begin{figure*}[!htb]
\begin{framed}
\small
\textbf{\em Network function transformation.} The client MB computes the tuple $(E(1), E(\mathbf{e}_i), E(y), E(\mathbf{e}_j), E(z\mathbf{e}_j))$ and sends it to the cloud MB. 

\medskip
\textbf{\em Packet processing.} Upon receiving a packet $\mathbf{x}$ the cloud MB:
		\setlist{nolistsep}
		\begin{enumerate}
		\itemsep0em 
		\item Encrypts the packet as $E(\mathbf{x})$. 
		\item Computes $E(a(\mathbf{x}))$ according to Eq.~\ref{eq:bgn-action} as $E(a(\mathbf{x})) = E(\mathbf{x}) - E(\mathbf{x}) \circ E(\mathbf{e}_j) + E(z\mathbf{e}_j)$ \\ 
		and $E(m(\mathbf{x})) = E(c)$ according to Eq.~\ref{eq:bgn-match} as $E(c) = E(1) - \langle E(\mathbf{x}), E(\mathbf{e}_i) \rangle + E(y)$
		\item Sends $E(\mathbf{x})$, $E(a(\mathbf{x}))$ and $E(c)$ to the client MB.
		\end{enumerate}
\medskip

\textbf{\em Packet decryption.} Upon receiving $E(\mathbf{x})$, $E(a(\mathbf{x}))$ and $E(c)$ the client MB:
	\begin{enumerate}
		\itemsep0em 
		\item Decrypts $E(c)$ to obtain $c$. 
		\item If $c = 1$, decrypts $E(a(\mathbf{x}))$ to obtain the transformed packet.
		\item Else if $c \ne 1$, decrypts $E(\mathbf{x})$ to obtain the unchanged packet. 
	\end{enumerate}
\end{framed}
\vspace{-0.4cm}
\caption{PNFV scheme based on the BGN cryptosystem~\cite{bgn}.}
\label{fig:bgnsol}
\end{figure*}

\descr{Range matching} Next, we consider range matching, i.e., the network function $\psi$ defined by policy~\ref{eq:range-pol}. Observe that: 
\begin{align*}
(b - x_i)(x_i - a) \ge 0 & \text{ ~~if } x_i \in [a, b] \\
(b - x_i) (x_i - a)< 0  & \text{ ~~otherwise}.
\end{align*}
The product above can be written as
\begin{align*}
(b - x_i)(x_i - a) &= bx_i - ab -x^2_i + ax_i \\
					   &= -x^2_i + (a + b)x_i -ab.
\end{align*}
Let $\mathbf{x}^2 = \langle \mathbf{x}, \mathbf{x} \rangle$. If we define the matching function as:
\[
 m(\mathbf{x}) = -\langle \mathbf{x}^2, \mathbf{e}_i \rangle + \langle \mathbf{x}, (a + b)\mathbf{e}_i \rangle - ab
\]
then $m(\mathbf{x}) \ge 0$ if there is a match, and negative otherwise. Homomorphically, we obtain:
\begin{align*}
E(m(\mathbf{x})) &= E(-\langle \mathbf{x}^2, \mathbf{e}_i \rangle + \langle \mathbf{x}, (a + b)\mathbf{e}_i \rangle - ab) \\
						& = -E(\langle \mathbf{x}^2, \mathbf{e}_i \rangle) + E(\langle \mathbf{x}, (a + b)\mathbf{e}_i \rangle) - E(ab)\\
						& = -\langle E(\mathbf{x}^2) , E(\mathbf{e}_i) \rangle + \langle E(\mathbf{x}), E((a + b)\mathbf{e}_i) \rangle - E(ab).
\end{align*}
Here, $E(ab)$, $E((a + b)\mathbf{e}_i)$ and $E(\mathbf{e}_i)$ are computed by the client MB during the setup phase as part of the network function transformation routine. Since the cloud MB already knows $\mathbf{x}$, it can compute $\mathbf{x}^2 = \langle \mathbf{x}, \mathbf{x} \rangle$ in the clear, and then compute $E(\mathbf{x}^2)$.  %
The action function is the same as before. The client MB  receives $E(\mathbf{x})$, $E(a(\mathbf{x}))$ and $E(m(\mathbf{x})) = E(c)$, and decrypts $E(c)$ to obtain $c$. If $c$ is a non-negative integer then it decrypts the result of the action function as the transformed packet, otherwise, it decrypts the original packet as the packet to be retained.

\descr{Correctness} As mentioned, BGN can successfully decrypt homomorphic encryptions of unlimited additions and one multiplication (per ciphertext). The above construction of match and action functions satisfy this constraint,
thus implying correctness of the PNFV scheme in Figure~\ref{fig:bgnsol}.

\descr{Privacy} Intuitively, the scheme can be shown to be private as the adversary only sees randomized encryptions of matching and action functions and, as such, cannot infer whether the matching function resulted in $1$ or some other value. More formally, we prove, in
Appendix~\ref{app:sec-red},
that, if BGN is semantically secure, our PNFV scheme is private against $\cal{A}_\text{strong}$.

\descr{Discussion} Ideally, the client MB would receive the encryption of the whole network function, i.e., $E(\psi(\mathbf{x}))$ and simply decrypt it to get the final packet. In our protocol, it actually has to perform two decryption operations instead of one (one to check the output of the matching function and another to decrypt the result), and, for each packet, three encryptions need to be sent. %
This is due to the fact that the output of the matching function is a variable (i.e., not a constant value) when there is no match, thus, we cannot perform iterations of $N$ network functions. 

\descr{Asymptotic Complexity} The network function transformation phase (which is done only once, during the 
setup) requires the client MB to compute, and send to the cloud MB, $O(N\cdot n)$ encryptions.
The packet processing at the cloud MB requires the computation of $O(N\cdot n)$ encryptions, which are then sent to the client MB.
Finally, the packet decryption at the client MB requires $O(N\cdot n)$ decryptions.

\subsection{Privacy against the Weak Adversary}
\label{sub:weak-peks}
We now present a more efficient solution that is secure against the weak adversary,
based on {\bf Public-key Encryption with Keyword Search (PEKS)}~\cite{peks}, a probabilistic encryption scheme $(E, D)$ and a pseudorandom permutation $\sigma$. %

Figure~\ref{fig:weakadvsol} presents our solution, in the context of policy~\ref{eq:equal-pol}.
Observe that $I$ denotes the $n$-element index vector whose $i$-th element is $i$ itself, and $\mathbf{x} || I$ the $n$-element vector whose $i$-th element is $x_i || i$. 
In this model, the weak adversary $\cal{A}_\text{weak}$ does not have access to the entry MB packet processing.
Thus, we have a somewhat stronger assumption of security in this scheme with respect to the strong adversary schemes presented in Section~\ref{sec:strongAdv}. The advantage, compared to the BGN based scheme presented in Section~\ref{sub:bgn-scheme}, is that we only send one encrypted packet, and the client MB only needs 
to decrypt the packet.
Also note that the entry MB runs $\sigma$ only once per packet arrival to obtain a shuffled set of indexes, and permutes the encryptions according to this set. That is, steps 1, 2 and 3 in Figure~\ref{fig:weakadvsol} performed by the entry MB use the same permutation.  %

\begin{figure*}[!htb]
\small
\begin{framed}
\textbf{\em Network function transformation.} Using PEKS, the client MB computes the trapdoors $T(y || i)$ and $T(j)$. Using $E$, the client MB creates the encryption $E(z || j)$. The client MB sends $T(y || i)$, $T(j)$ and $E(z || j)$ to the cloud MB.

\medskip
\textbf{\em Packet processing.} This is divided into entry MB and cloud MB.

\medskip
\textit{\em Entry MB}: Upon receiving a packet $\mathbf{x}$:
		\setlist{nolistsep}
		\begin{enumerate}
		\itemsep0em 
		\item Encrypts $\mathbf{x} || I$ using $E$ and shuffles the result as $\sigma(E(\mathbf{x}|| I))$.
		\item Encrypts $\mathbf{x} || I$ using PEKS and shuffles the result as $\sigma(\mathcal{E}(\mathbf{x}|| I))$.
		\item Encrypts $I$ using PEKS and shuffles it as $\sigma(\mathcal{E}(I))$.
		\item Deletes the original packet $\mathbf{x}$.
		\end{enumerate}

\medskip
\textit{\em Cloud MB}: Upon receiving $\sigma(E(\mathbf{x}|| I))$, $\sigma(\mathcal{E}(\mathbf{x}|| I))$ and $\sigma(\mathcal{E}(I))$:
		\begin{enumerate}
		\itemsep0em 
		\item Checks if there exists an $l \in [n]$ such that $\mathsf{test}( \mathcal{E}( x_l || l ), T(y || i )) = 1$. 
			\begin{enumerate}[label={\arabic{enumi}.\arabic*.}]
			\itemsep0em 
			\item If yes, finds an $l' \in [n]$ such that $\mathsf{test}( \mathcal{E}( l' ) , T(j) ) = 1$ (which should exist). 
			\item Replaces $E(x_{l'} || l')$ with $E(z || j)$ in $\sigma(E(\mathbf{x}|| I))$ and sends it to the client MB.			
			\end{enumerate}				
		\item Else, sends $\sigma(E(\mathbf{x}|| I))$ to the client MB.
		\end{enumerate}
		
\medskip
\textbf{\em Packet decryption.} The client MB upon receiving $\sigma(E(\mathbf{x}|| I))$, decrypts to obtain $\sigma(\mathbf{x} || I)$ and then reconstructs $\mathbf{x}$ according to $I$.
\end{framed}
\vspace{-0.4cm}
\caption{Scheme based on PEKS, private against the weak adversary.}
\label{fig:weakadvsol}
\end{figure*}

\descr{Correctness} The client MB decrypts $E(\mathbf{x}'|| I)$, permuted by $\sigma$, to obtain $\mathbf{x}' || I$ and  reconstructs $\mathbf{x}'$ according to $I$. Note that, if the original packet matches policy~\ref{eq:equal-pol}, then $x'_j = z$. Likewise, it the packet does not match the policy, decrypted packet $\mathbf{x}'$ is the original packet $\mathbf{x}$. Therefore, our PNFV scheme is correct.

\descr{Privacy} Intuitively, since $\cal{A}_\text{weak}$ does not know which packet index yields a match and which index the action applies to (due to random shuffle by $\sigma$), and since the matching value $y$ and the action value $z$ are encrypted, it cannot infer the policy. In 
Appendix~\ref{app:sec-red},
we show that if the probabilistic encryption scheme $E$ is semantically secure and the PEKS scheme is semantically secure against a chosen keyword attack, its trapdoor function $T$ is not invertible, and the pseudorandom permutation $\sigma$ is indistinguishable from a random permutation, then the PNFV scheme described in Figure~\ref{fig:weakadvsol} is private against $\cal{A}_\text{weak}$.

\descr{Discussion} The obvious limitation of this scheme is that it is only private against a weaker notion of adversary. In particular, we consider a cloud MB that does not try to analyze incoming packet $\mathbf{x}$ with the output of the scheme. More precisely, the cloud MB does not retain the packet $\mathbf{x}$ to match its randomly permuted encryptions, neither does it attempt to find $j$ in $T(j)$ by checking all possible encryptions under $\mathcal{E}$ of all possible elements in $[n]$. If the cloud MB tries to do either of these (unwarranted) actions, it will at best learn the index $j$ (and not index $i$, $y$ or $z$). To find $(i, y)$, the cloud MB needs to do a brute force search whose complexity is $O(2^{qn})$. %
On the other hand, we only need to send a number of encryptions per packet independent	of the number of network functions $N$, however, it is only applicable to policy~\ref{eq:equal-pol}.

\descr{Asymptotic Complexity}
The network function transformation, similarly to the BGN based scheme, is done only
once, during the setup, by the client MB, computing $O(N)$ PEKS trapdoors and encryptions to be sent to the cloud MB.
The packet processing requires the entry MB to perform and send to the cloud MB $O(N)$ encryptions,
while the cloud MB performs $O(N\cdot n)$ equality tests (using the PEKS scheme), for each packet, and sends $O(n)$ ciphertexts to the client MB. Finally, the client MB needs to decrypt $O(n)$ ciphertexts.

\subsection{Handling State Tables}
We now discuss PNFV solutions in the context of stateful network functions. Recall that a stateful network function maintains a state table, which among others contains a field (column) labelled \emph{state}. We model a state table as comprising of one or more packet field headers followed by a \emph{state} field and an \emph{action tag}. 

The private state table solution is built from the PEKS based PNFV scheme discussed above.
Note that FHE based solutions are not applicable to state tables, as the cloud MB should discontinue processing once a match is found in the state table. If processing needs to be continued for the packet, and the current state table only maintains statistics (such as counters), then this can be implemented in the same way as a normal network function. We denote the state and tag fields by $s$ and $t$ respectively. 

Our proposed solution is shown in Figure~\ref{fig:statetablesol}. In case no entry in the state table is found, the cloud MB processes the static network policies via the underlying PNFV scheme. In Figure~\ref{fig:statetablesol}, we assume that this is the BGN based PNFV scheme from Section~\ref{sub:bgn-scheme}. If relying on the PEKS-based scheme, the entry MB needs to send $\sigma(E(\mathbf{x}|| I))$ and $\sigma(\mathcal{E}(I))$ to the cloud MB instead of $\mathbf{x}$.

\begin{figure*}[!htb]
\small
\begin{framed}
{Client MB}: Upon receiving a packet $\mathbf{x}$ from the cloud MB decides that a state table entry is to be created. 
		\setlist{nolistsep}
		\begin{enumerate}
		\itemsep0em 
		\item Identifies a subset $I'$ of $I$ corresponding to packet fields to be placed in the state table.
		\item Produces trapdoors $T(\mathbf{x}_{I'} || I')$ and shuffles them using $\sigma$ as $\mathcal{T} = \sigma(T(\mathbf{x}_{I'} || I'))$. 
		\item Creates encryptions of the state and the tag as $E(s)$ and $E(t)$, respectively.
		\item Sends $\mathcal{T} = \sigma(T(\mathbf{x}_{I'} || I'))$, $E(s)$ and $E(t)$ to the cloud MB. 
		\end{enumerate}		

\medskip
{Cloud MB}: Creates a state table entry with $\mathcal{T} = \sigma(T(\mathbf{x}_{I'} || I'))$, $E(s)$ and $E(t)$, and sends the $\texttt{id}$ of this entry to the client MB.

\medskip
{Entry MB}: Upon receiving a packet $\mathbf{x}$, encrypts $\mathbf{x} || I$ using PEKS and shuffles the result as $\sigma(\mathcal{E}(\mathbf{x}|| I))$.

\medskip
{Cloud MB}: Upon receiving $\mathbf{x}$ and $\sigma(\mathcal{E}(\mathbf{x}|| I))$, for $l' \in |\mathcal{T}|$ checks whether there exists an $l \in [n]$ such that $\mathsf{test}( \mathcal{E}( x_l || l ), T( x_{l'} || l' )) = 1$. 
		\begin{enumerate}
			\itemsep0em 
			\item If there is a match for all $l'$, computes $E(\mathbf{x})$, appends $E(\texttt{id})$, $E(s)$ and $E(t)$ to it and sends it to the client MB.
			\item Otherwise, continues processing the \emph{static} network functions (using the PNFV scheme).
		\end{enumerate}				

\medskip
{Client MB}: Upon receiving an encrypted packet $E(\mathbf{x})$
		\begin{enumerate}
		\itemsep0em 
		\item Decrypts it to obtain $\mathbf{x}$.
		\item Strips the $\texttt{id}$, state $s$ and tag $t$, and carries out the action according to $t$.
		\item \textit{Update:} Sends the tuple $(\texttt{id}, E(s'))$ to the cloud MB, where $s'$ is the new state.
		\item \textit{Deletion:} Sends the tuple $(\texttt{id}, \texttt{delete})$ to the cloud MB.
		\end{enumerate}
\end{framed}
\vspace{-0.4cm}
\caption{State table solution private against the weak adversary. The PNFV scheme used in case of a state table miss is based on BGN.}
\label{fig:statetablesol}

\end{figure*}

\descr{Example} 
We illustrate the state table solution using a firewall state table as an example. The client MB identifies the index set 
\begin{equation}
\label{eq:5tuple}
I' = \{ \texttt{s\_ip}, \texttt{d\_ip}, \texttt{s\_port}, \texttt{d\_port}, \texttt{prot}\}, 
\end{equation}
which correspond to the source IP address, destination IP address, source port, destination port and protocol fields, respectively, of an IPv4 packet. The client MB creates trapdoors for the values of these fields and randomly shuffles 
the trapdoors creating the set $\cal {T}$. Let $s \in \{\texttt{new}, \texttt{est}\}$ be the possible states, where the second state is the abbreviation of ``established.'' Let $t \in \{\texttt{allow}, \texttt{drop} \}$ be the possible tags. Suppose the client MB first receives a packet $\mathbf{x}$ whose SYN flag is set. The client MB then sends the set $\cal T$ and $E(s) = E(\texttt{new})$ and $E(t) = E(\texttt{allow})$ to the cloud MB. The cloud MB creates an entry for this state table. Suppose the identifier of this state table entry is $\texttt{id}$, which is sent to the client MB. The cloud MB subsequently checks each incoming encrypted and shuffled packet $\mathbf{x}$ (done by the entry MB) to see if it matches this state table entry. If it does, it simply appends $E(\texttt{id})$, $E(\texttt{new})$ and $E(\texttt{allow})$ as the state and tag respectively, to the encrypted packet $E(\mathbf{x})$ and sends it to the client MB. The client MB, after decrypting the packet, checks the state and the ACK flag in $\mathbf{x}$. If the ACK flag is set, the client MB sets $s \leftarrow \texttt{est}$ and sends $(\texttt{id}, E(\texttt{est}))$ to the cloud MB. Since the tag is set to $\texttt{allow}$, it forwards the packet to 
its intended destination within the internal network. After the current TCP connection is over (through FIN-ACK exchange), the client MB sends the pair $(\texttt{id}, \texttt{delete})$ to the cloud MB, which in turn deletes the corresponding entry.

\descr{Privacy} The privacy argument of the proposed state table solution is similar to the one for the PEKS based PNFV scheme, and hence we omit it here. However, two important differences are that (i) the adversary $\mathcal{A}_\text{weak}$ knows the number of fields being checked (due to $|I'|$), and (ii) learns whether or not the current packet matches a state table entry. %

\section{Implementation and Performance Evaluation}
\label{sec:implement}
In the following, we provide a proof-of-concept of the feasibility of our PNFV schemes. 

\subsection{Implementing PNFV}
We assume that private network function processing operates at the network layer in the OSI model, i.e., it processes IP packets, although it can be extended to the processing of Ethernet frames as well (MAC headers). 
Note that not all packet fields are needed for private processing of a given network function, e.g., the ``header checksum'' field of an IPv4 packet is used for integrity check and does not have to be encrypted. Thus, we only use a subset $I'$ of the set of indexes $I$ corresponding to different fields of a packet. For instance, if the network function performs firewall actions, we assume that $I'$ is the 5-tuple defined in Eq.~\ref{eq:5tuple}. %
The packet encryption algorithm of the cloud MB, upon receiving the packet $\mathbf{x}$, computes encryptions of the above fields only. Recall that, in the PEKS based scheme, the entry MB deletes the original packet $\mathbf{x}$: for this tuple, this implemented by the entry MB resetting the corresponding field values to $0$ before sending $\mathbf{x}$ to the cloud MB.

Whenever a packet $\mathbf{x}$ arrives at the cloud MB, after private processing, this is transformed into a new packet $\mathbf{x}'$ (as shown in Figure~\ref{fig:encapsulation-x}) which is then sent to the client MB. For instance, assume a network function implementing the policy: if $x_{\texttt{s\_ip}} = \texttt{127.0.0.1}$ then block the packet, otherwise allow it, and %
assume we are using the PNFV scheme based on BGN (Section~\ref{sub:bgn-scheme}). 
The client MB constructs the transformed packet $\mathbf{x}'$ as follows. It first constructs a new IP header containing its IP address as the source IP and the IP address of the client MB as the destination IP (similarly for the ports). The payload of $\mathbf{x}'$ contains the original IP packet $\mathbf{x}$, as shown in the figure, followed by PNFV related payload. The PNFV payload specific to the BGN based scheme and the above mentioned policy is:

\begin{figure}[!tb]
\begingroup%
  \makeatletter%
  \providecommand\color[2][]{%
    \errmessage{(Inkscape) Color is used for the text in Inkscape, but the package 'color.sty' is not loaded}%
    \renewcommand\color[2][]{}%
  }%
  \providecommand\transparent[1]{%
    \errmessage{(Inkscape) Transparency is used (non-zero) for the text in Inkscape, but the package 'transparent.sty' is not loaded}%
    \renewcommand\transparent[1]{}%
  }%
  \providecommand\rotatebox[2]{#2}%
  \ifx\svgwidth\undefined%
    \setlength{\unitlength}{228.60229492bp}%
    \ifx\svgscale\undefined%
      \relax%
    \else%
      \setlength{\unitlength}{\unitlength * \real{\svgscale}}%
    \fi%
  \else%
    \setlength{\unitlength}{\svgwidth}%
  \fi%
  \global\let\svgwidth\undefined%
  \global\let\svgscale\undefined%
  \makeatother%
  \begin{picture}(1,0.44079526)%
    \put(0,0){\includegraphics[width=\unitlength]{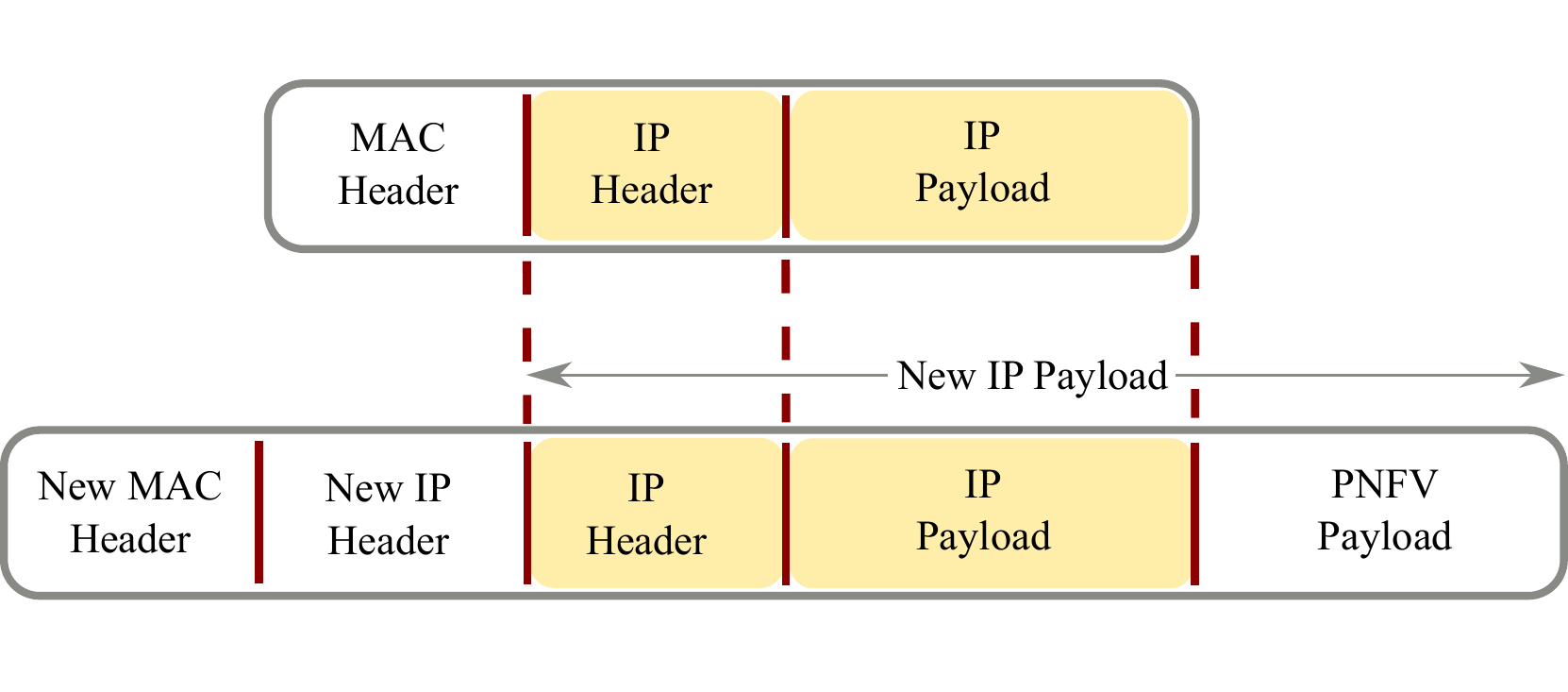}}%
    \put(0.270219743,0.42047569){\color[rgb]{0,0,0}\makebox(0,0)[lb]{\smash{\small Ethernet frame of packet $\mathbf{x}$}}}%
    \put(0.27606589,0.00556308){\color[rgb]{0,0,0}\makebox(0,0)[lb]{\smash{\small Ethernet frame of packet $\mathbf{x}'$}}}%
  \end{picture}%
\endgroup%
\caption{Encapsulation of packet $\mathbf{x}$ within packet $\mathbf{x}'$ by the cloud MB. Fields shaded \textcolor{nfvyellow}{\protect\rule[0pt]{2mm}{2mm}}~have identical content.}
\label{fig:encapsulation-x}
\vspace{-0.3cm}
\end{figure}

\medskip\noindent
\resizebox{\columnwidth}{!}{
\begin{tabular}{ccccc} 
\text{PNFV ID} & \text{state/policy ID} & $x_j || j$ & $z || j$ & $c$ \\
\hline
\texttt{BGN} & \texttt{id} & $E(\texttt{allow} || \texttt{tag})$ & $E(\texttt{deny}  || \texttt{tag})$ & $E(c)$ 
\end{tabular}
}
\medskip

The first field contains the ID of the PNFV scheme being used (in this case, the BGN based scheme). The second field reports the policy ID of the particular policy being processed (in case of state tables this is the state table entry ID). The next two items are the two possible actions that are to be applied on packet field $j$ depending on whether there was a policy match. In our example, this is $E(\texttt{allow} || \texttt{tag})$ which is the action when there is no match and $E(\texttt{deny} || \texttt{tag})$ is the action when there is a match. Here $\texttt{tag}$ is an index for the \emph{virtual} field tag, since IP packets do not have a $\texttt{tag}$ field. In case of other policies, this could be a real packet field, for instance the $\texttt{prot}$ (protocol) field. The last item is the encryption of the result of the matching function, i.e., $E(c)$.

When the client MB receives the packet $\mathbf{x}'$, it first extracts $\mathbf{x}$ in a straightforward manner. The client MB then checks the PNFV ID to learn which scheme is to be applied (in this case, the BGN based scheme) and decrypts the last item $E(c)$. If $c = 1$, it decrypts $E(\texttt{deny} || \texttt{tag})$ and then drops the packets $\mathbf{x}$,
whereas, if $c \ne 1$, it decrypts $E(\texttt{allow} || \texttt{tag})$, and forwards the packet $\mathbf{x}$ to its intended destination. The policy ID can be used for bookkeeping.  
If the original packet $\mathbf{x}$ has size $| \mathbf{x} |$, then the size of $\mathbf{x}'$ is given by:
\[
|\mathbf{x}' | = | \mathbf{x} | + \text{New IP header} + \text{PNFV payload}
\]
As an example, consider the smallest sized packet $\mathbf{x}$ of 34 bytes (14 bytes for the MAC header, 20 bytes for the IP header and 0 bytes for the payload). %
If PNFV ID requires 4 bits and the state/policy ID requires another 20 bits, and the BGN ciphertexts have a blocksize of $256$ bits, then the PNFV payload has 99 bytes, thus yielding $34 + 20 + 99 = 153$ bytes for $\mathbf{x}'$. In the case of the PEKS based scheme, the overhead is actually higher since encryptions corresponding to the 5-tuples and the virtual $\texttt{tag}$ field needs to be added, thus yielding a tradeoff between packet processing efficiency and bandwidth/storage overhead.

\subsection{Empirical Evaluation}
We implemented the PEKS based and BGN based schemes in C, using the \texttt{RELIC} cryptographic library~\cite{relic-toolkit}. As discussed earlier, the PEKS based scheme relies on the Boneh and Franklin cryptosystem~\cite{ibe}, whereas, for the BGN based scheme, we modified the Freeman's prime-order version~\cite{freeman2010converting} provided by \texttt{RELIC} in order to fix some bugs in the decryption phase and to implement lookup tables of pre-computed discrete logarithms in order to achieve constant-time decryption. For the two schemes, we chose a Barreto-Naehrig pairing-friendly elliptic curve defined on a 256-bit prime order group, achieving a 128-bit security level. For pairing computations, we used the optimal ate pairing implementation provided by \texttt{RELIC}.

In the following, we present empirical results on PNFV simulations using the generic policy~\ref{eq:equal-pol}, setting the size of each packet attribute $x_i$ to 4 bytes, which is the largest size of an IP header field in IPv4 packets (corresponding to IP addresses).  Simulations were performed on a machine running Ubuntu Trusty Tahr (Ubuntu 14.04.2 LTS), equipped with a 2.4 GHz CPU i5-520M and 4GB RAM.

\begin{figure*}[!htb]
\centering
    \begin{subfigure}[t]{0.3\textwidth}	\captionsetup{skip=0pt}
        \centering
		\includegraphics[width=\columnwidth]{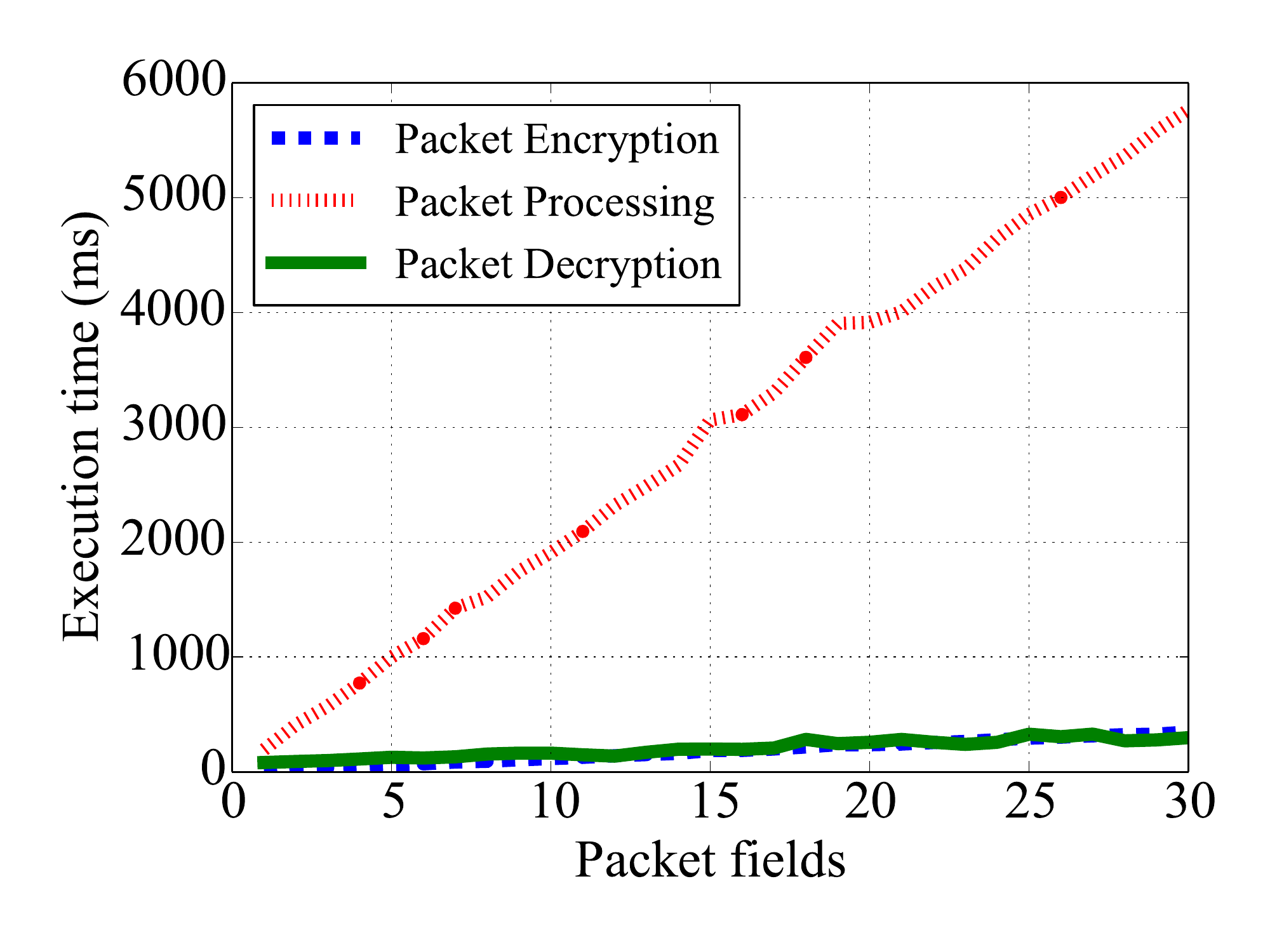}
        \caption{BGN based scheme with 10 policies\label{fig:time_strong_attr}}
    \end{subfigure}%
~
\begin{subfigure}[t]{0.3\textwidth} \captionsetup{skip=0pt}
        \centering
		\includegraphics[width=\columnwidth]{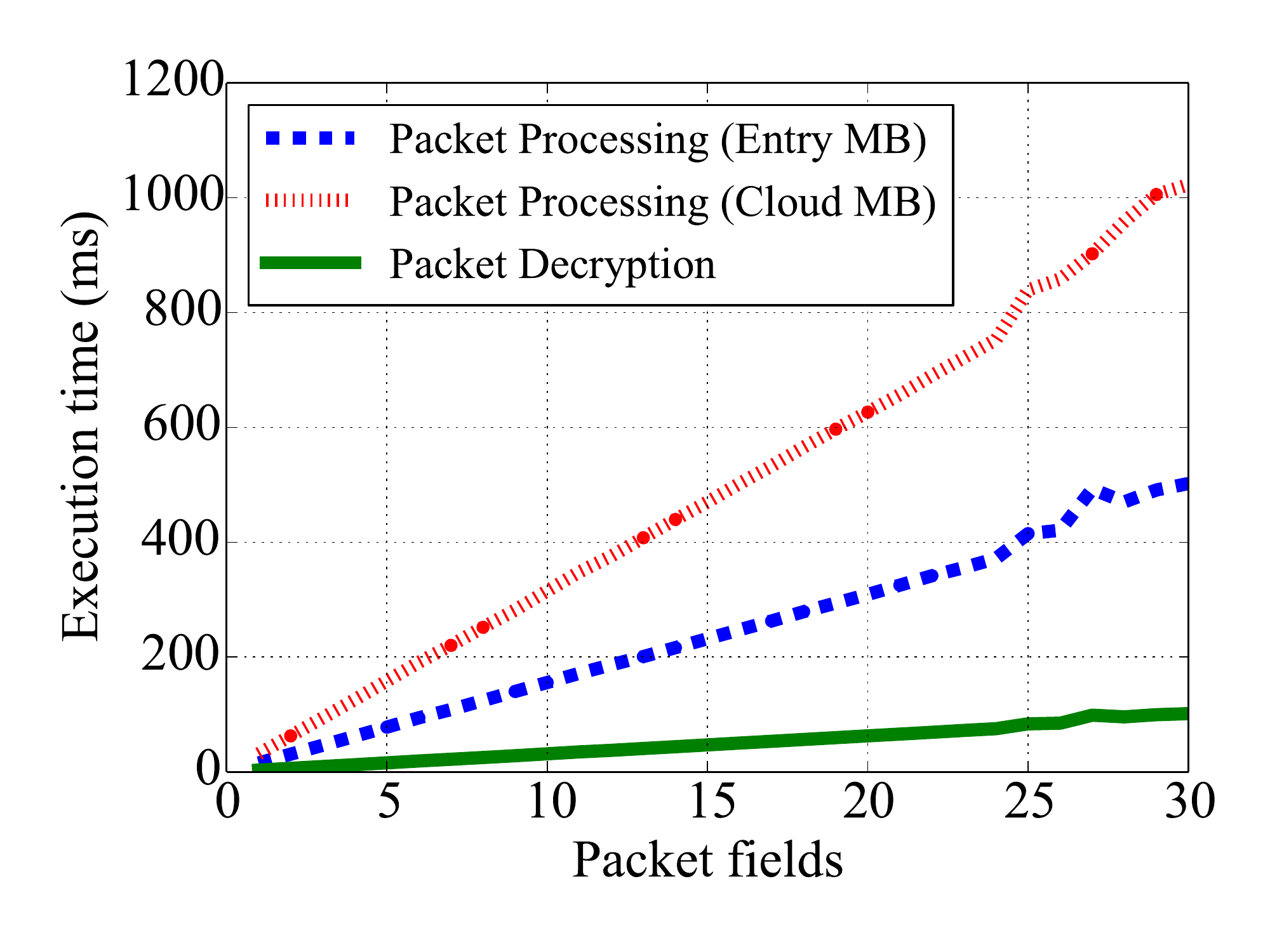}
        \caption{PEKS based scheme with 10 policies\label{fig:time_weak_attr}}
    \end{subfigure} 
~
	\begin{subfigure}[t]{0.3\textwidth} \captionsetup{skip=0pt}
        \centering
		\includegraphics[width=\columnwidth]{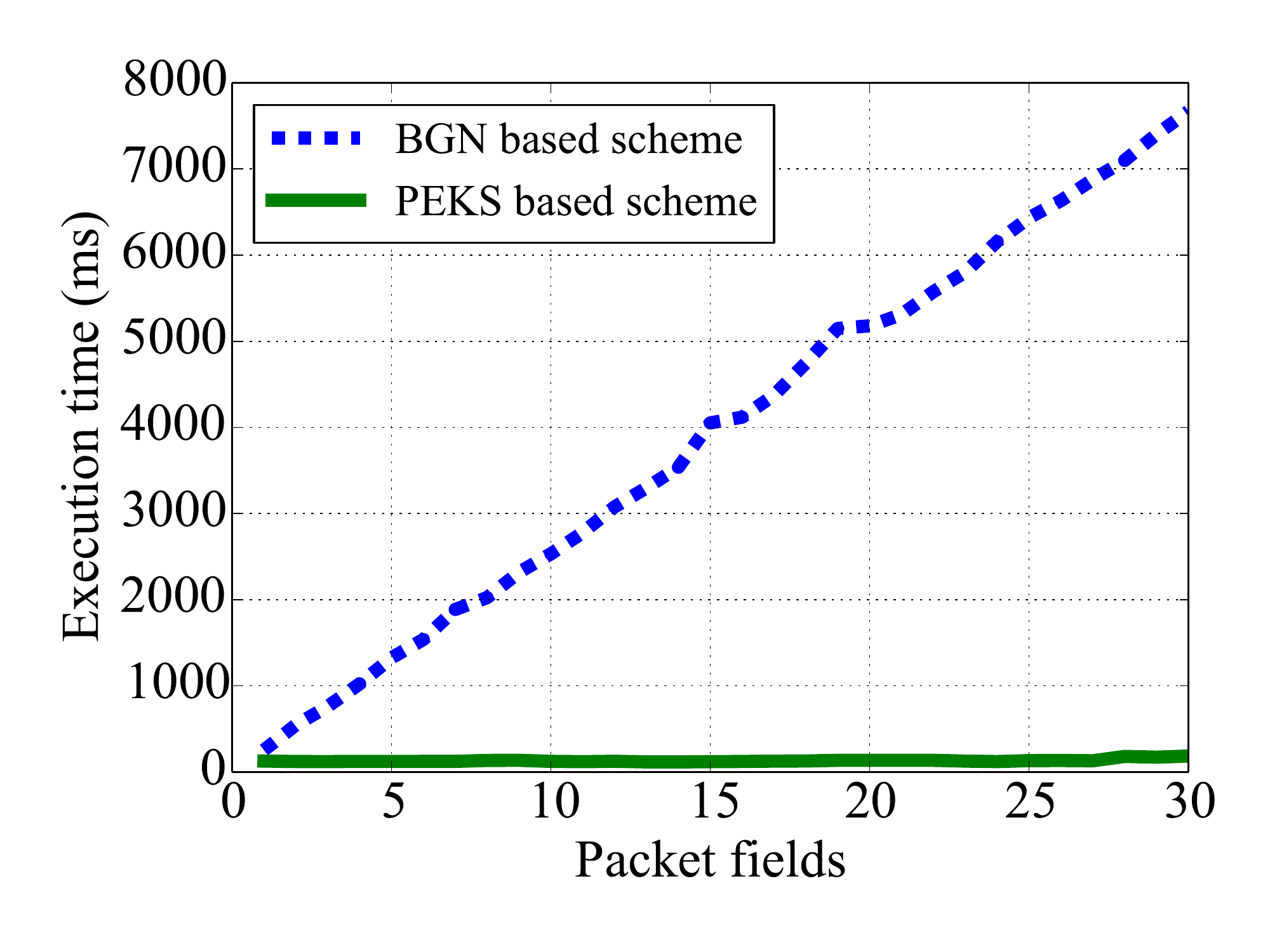}
        \caption{Network function transformation with 10 policies (setup) \label{fig:clientMB_attributes}}
\vspace*{-0.1cm} 
    \end{subfigure} 
~
    \begin{subfigure}[t]{0.3\textwidth} \captionsetup{skip=0pt}
        \centering
		\includegraphics[width=\columnwidth]{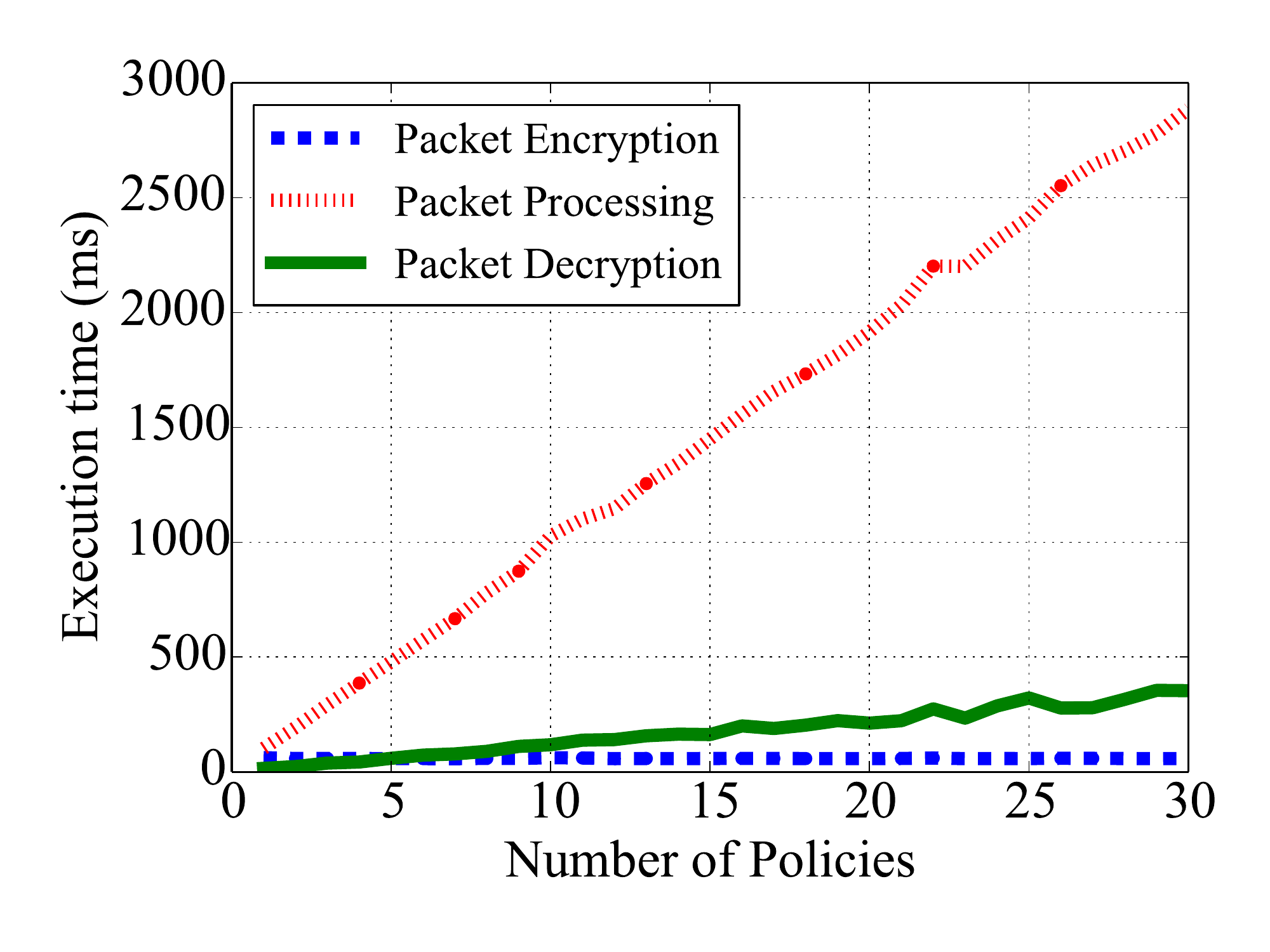}
        \caption{BGN based scheme with 5 packet fields\label{fig:time_strong_policies} }
    \end{subfigure} 
~
     \begin{subfigure}[t]{0.3\textwidth} \captionsetup{skip=0pt}
        \centering
		\includegraphics[width=\columnwidth]{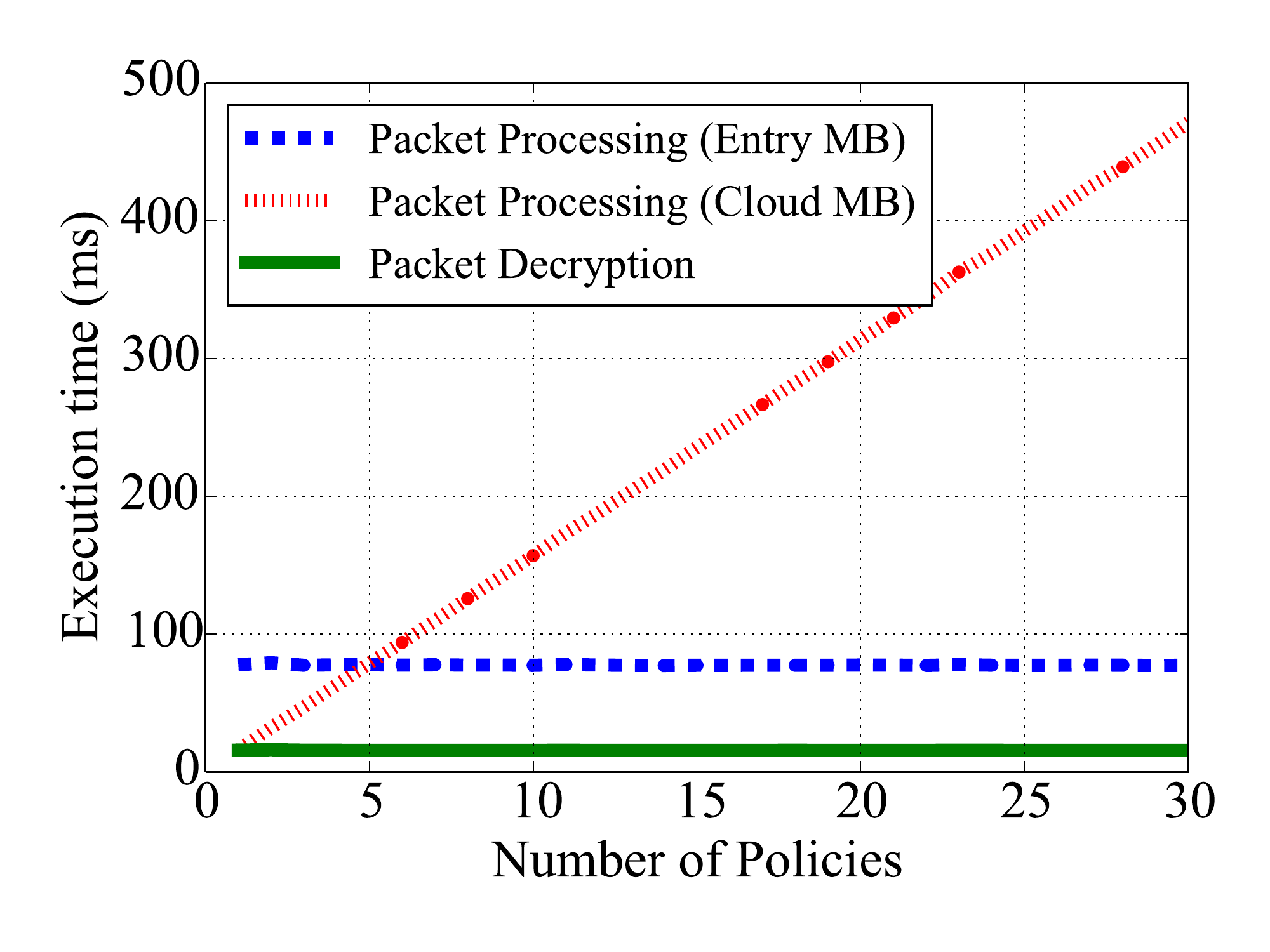}
        \caption{PEKS based scheme with 5 packet fields\label{fig:time_weak_policies} }
    \end{subfigure} 
~
	 \begin{subfigure}[t]{0.3\textwidth} \captionsetup{skip=0pt}
        \centering
		\includegraphics[width=\columnwidth]{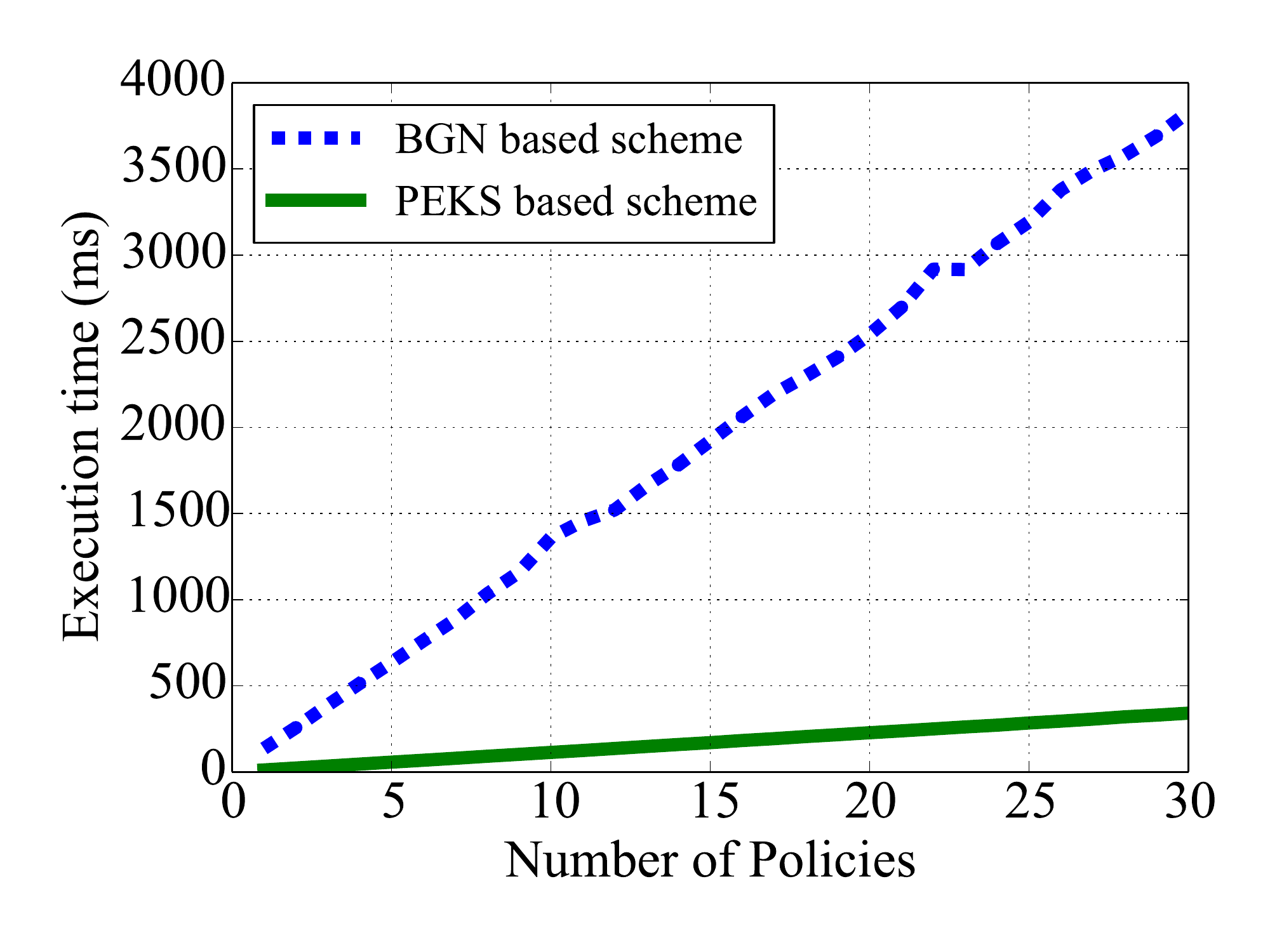}
        \caption{Network function transformation with 5 packet fields (setup)\label{fig:clientMB_policies} }
    \end{subfigure}     
  \vspace{-0.2cm}
\caption{\label{fig:time_strong} Execution times of different algorithms of the BGN and PEKS based schemes as functions of packet fields and number of policies.}
\end{figure*}

\descr{BGN based Scheme} Figures~\ref{fig:time_strong_attr} and~\ref{fig:time_strong_policies} report execution times of packet encryption, processing and decryption of the BGN based scheme w.r.t., respectively, the number of packet fields (and $10$ policies) and the number of policies (and $5$ packet fields). Experiments in Figure~\ref{fig:time_strong_policies} are intended to simulate a typical firewall rule that uses the $5$-tuple given by Eq.~\ref{eq:5tuple}. 

Note that the execution time of all three algorithms is linear in the number of packet fields (Figure~\ref{fig:time_strong_attr}). Whereas, as shown in Figure~\ref{fig:time_strong_policies}, execution times of packet processing and decryption are linear in the number of policies, but constant for packet encryption. For a network function with $10$ policies, private processing of $5$ packet fields takes 62 ms for encryption, 1,027 ms for processing, and 118 ms for decryption. 

Then, Figures~\ref{fig:clientMB_attributes} and \ref{fig:clientMB_policies} plot the execution time for the network function transformation algorithm: for the BGN based scheme, this is linear both as a function of the number of packet fields and policies, reaching a maximum of 7,669 ms (30 fields and 10 policies) and 3,831 ms (5 fields and 30 policies). However, note that these times are acceptable since this  does not have to be executed in real-time but only once, during the setup.

\descr{PEKS based Scheme} In Figures~\ref{fig:time_weak_attr} and~\ref{fig:time_weak_policies}, we report the execution times of the packet processing and decryption algorithms for the PEKS based scheme as a function of packet fields and number of policies. As the entry MB performs packet encryption and some preliminary packet processing, we divide the corresponding times between entry MB and cloud MB. 

Note from Figure~\ref{fig:time_weak_attr} that packet processing (both at entry MB and cloud MB) as well as decryption 
are linear w.r.t. increasing number of packet fields, while packet processing at entry MB and decryption are constant w.r.t. increasing number of policies (Figure~\ref{fig:time_weak_policies}). For a network function with 10 policies, private processing of $5$ packet fields takes 77 ms at the entry MB, 157 ms at the cloud MB and 16 ms for decryption. 

Finally, Figures~\ref{fig:clientMB_attributes} and \ref{fig:clientMB_policies} show that the network function transformation algorithm for this scheme is linear both in the number of packet fields and policies, reaching a maximum of 341 ms and 184 ms, respectively.

\descr{Comparison of the two schemes} Figures~\ref{fig:clientMB_attributes} and~\ref{fig:clientMB_policies} show the aggregate times of the two schemes (by adding up the times of packet encryption, processing and decryption) against increasing number of fields (with $10$ policies) and increasing number of policies (with $5$ packet fields used for private processing). The PEKS based scheme clearly outperforms the BGN based scheme. For instance, for a network function with $10$ policies, private processing of $5$ packet fields takes 250 ms in the PEKS based scheme and 1,208 ms in the BGN based scheme. 

\begin{figure}[!htb]
\centering
    \begin{subfigure}[t]{0.26\textwidth}
\hspace{-1cm}
	\captionsetup{skip=0pt}
        \centering
		\includegraphics[width=\columnwidth]{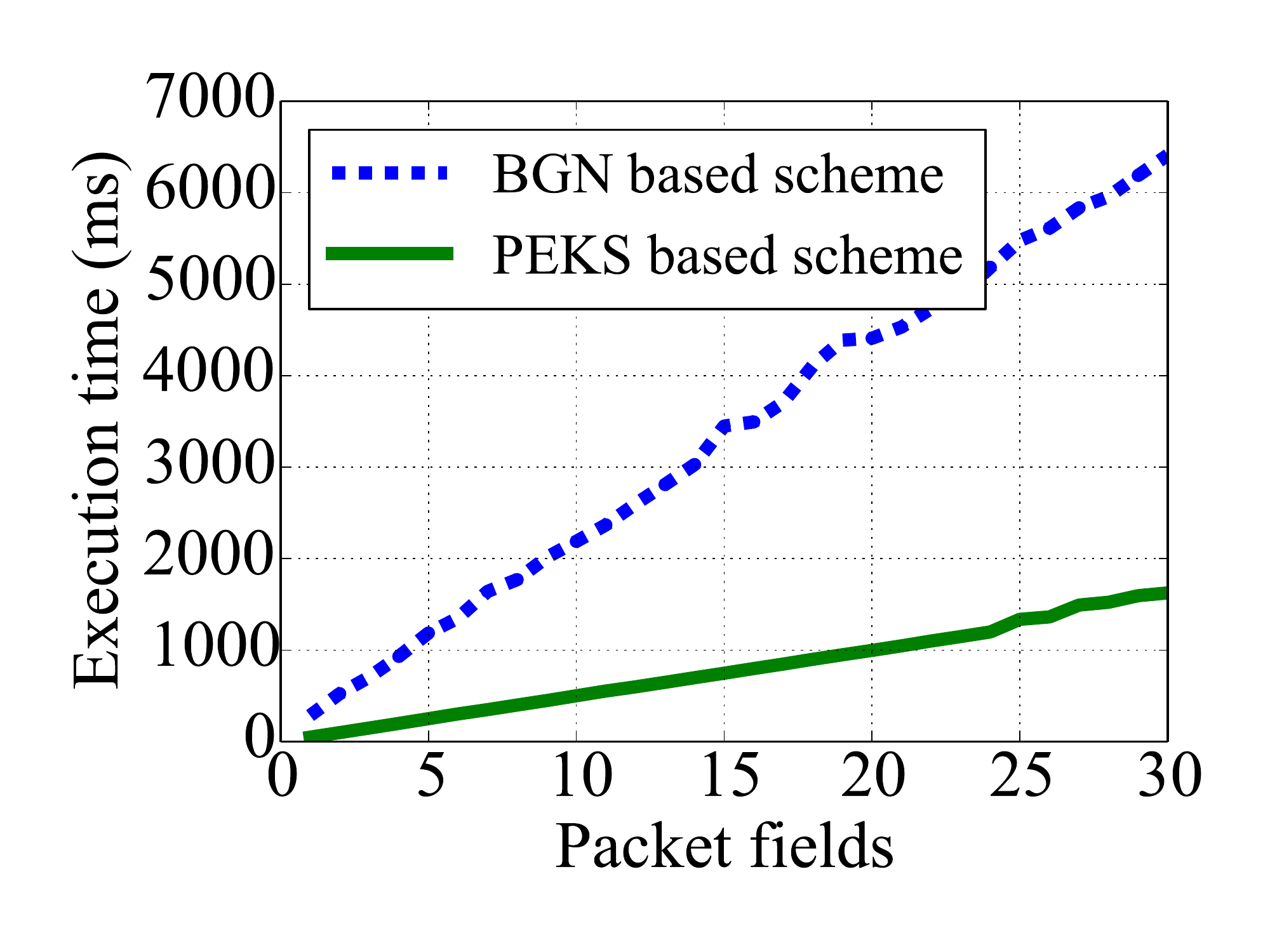}
        \caption{With 10 policies\label{fig:comparison_attributes} }
    \end{subfigure}%
    \hspace*{-1cm}
~
    \begin{subfigure}[t]{0.26\textwidth}
    	\captionsetup{skip=0pt}
        \centering
		\includegraphics[width=\columnwidth]{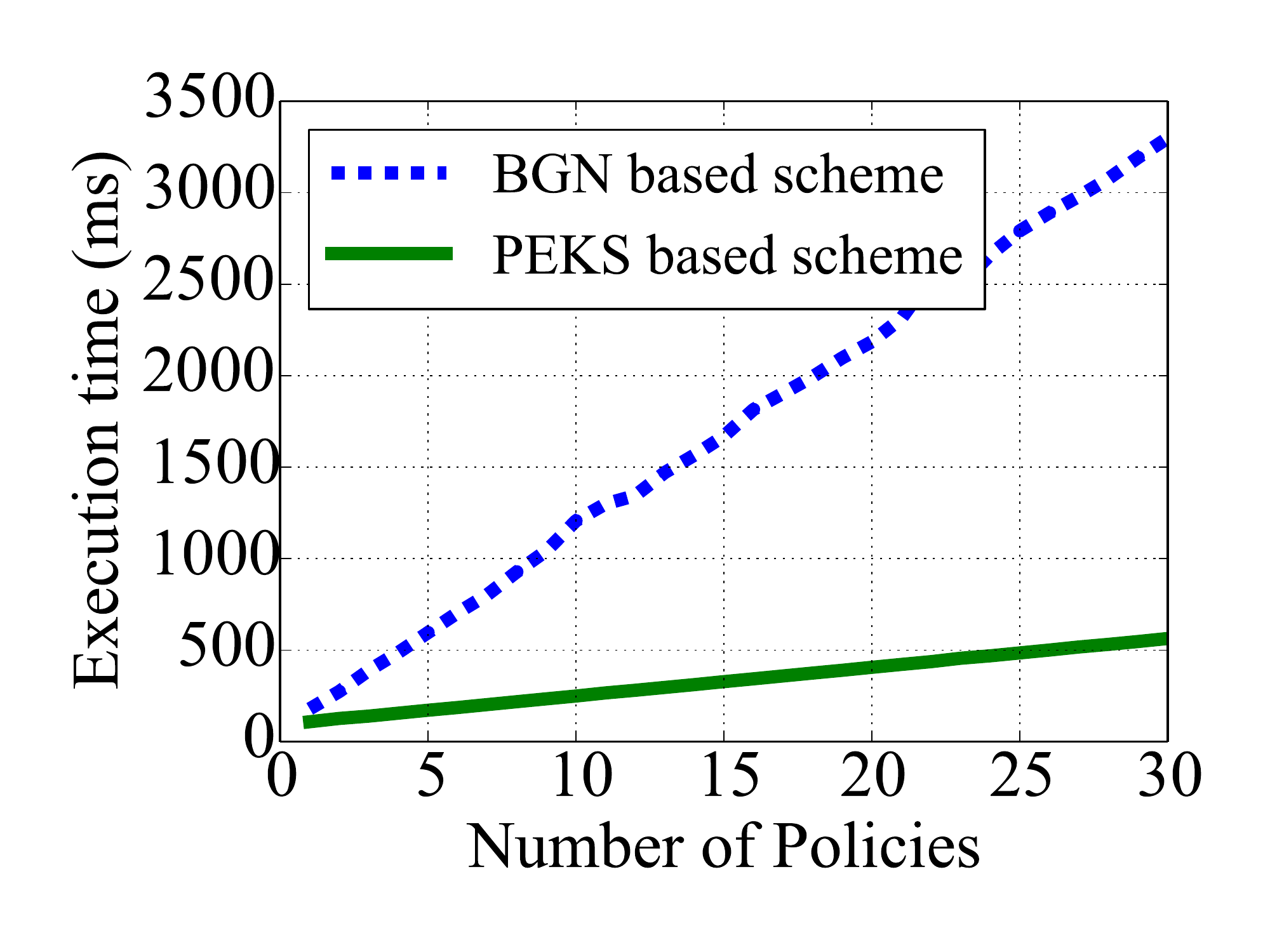}
        \caption{With 5 packet fields\label{fig:comparison_policies} }
    \end{subfigure} 
    \vspace{-0.2cm}
\caption{\label{fig:client_comparison} Aggregate execution times (packet encryption, processing and decryption) for the two schemes.}
    \vspace{-0.2cm}
\end{figure}

Translated into packets per second (pps), the above two numbers translate to a modest 4 pps and 0.82 pps, respectively. However, we remark that our implementation merely stands as a proof-of-concept, and as such we did not go for further implementation efficiency by using a more powerful machine or multi-threading in C. For instance, the time taken by the entry MB, the cloud MB and the client MB for a packet with a single encrypted field and a network function with a single policy was 13.32 ms, 5.41 ms and 2.69 ms, respectively, giving a total of 21.42 ms. Using multi-threading we can process a larger number of packet fields (in the case of the entry and client MB) and the policies (in the case of the cloud MB) in parallel, thus increasing the number of packets processed per second. With a modestly more powerful machine that can process say 50 threads concurrently, we can achieve a rate of more than 2,300 pps (using 21.42 ms as the baseline). 

Even without optimizations, e.g., multi-threading, our performance is comparable to that of the schemes proposed by Shi, Zhang and Zhong~\cite{mlm-firewall}. The three different \emph{modes} in~\cite{mlm-firewall} yield 60 ms, 1,000 ms and 3,000 ms for private processing of a $5$-tuple with $10$ firewall rules. The Bloom filter based scheme from Khakpour and Liu~\cite{bf-firewall} does much better, achieving 0.1 ms for a 10 rule firewall.\footnote{These approximate numbers are deduced from ACL index 16 from Figure~8 in~\cite{bf-firewall}.} However, as described in Section~\ref{sec:related}, both these works are narrower in scope and their security, at best, is questionable. 
\section{Conclusion}  
\label{sec:conclude}
This paper addressed the problem of private processing of outsourced network functions, where network function policies need to be kept private from the cloud, other tenants and third parties. 
We presented a cryptographic treatment of the problem, introducing security definitions as well as an abstract model of generic network functions, and proposed a few instantiations using homomorphic encryption and public-key encryption with keyword search. 
The performance of our proposed solutions is reasonable considering that we rely on public key operations and provide provable security in the presence of an honest-but-curious cloud, while guaranteeing that third party users, who are sending/receiving traffic, are oblivious to network function outsourcing. 
In future work, we plan to investigate mechanisms to further speed up computation, e.g., assuming that part of the cloud runs on a trusted computing base. %
We are also working on integrating our solutions for private NFV to existing NFV frameworks such as OPNFV\footnote{\url{https://www.opnfv.org/}.} and ClickOS~\cite{click-os}.

\appendix
\section{Proofs} 
\label{app:sec-red}
The following reductionist arguments will use policy~\ref{eq:equal-pol} as the base. It is straightforward to extend the arguments to policy~\ref{eq:range-pol}.
\begin{claim}
If the FHE scheme $(E, D)$ is semantically secure (indistinguishable under chosen plaintext attack), then the PNFV scheme based on it is private against $\cal{A}_\text{strong}$.
\end{claim}

\begin{argument}
We assume an FHE oracle which when given a plaintext $x$ returns the encryption $E(x)$, and when given two ciphertexts $E(x)$ and $E(y)$ returns $E(x + y)$. We use $\cal{A}_\text{strong}$ as a subroutine to an adversary $\cal{B}$ that tries to subvert the FHE scheme. $\cal{B}$ announces $m_0 = z_0$ and $m_1 = z_1$ as its chosen plaintexts. $\cal{B}$ is given $E(m_b)$ such that $b = 0$ with probability $\frac{1}{2}$, and is asked to guess $b$. 

$\cal{B}$ begins by choosing a $y \ne m_0, m_1$ and requesting the encryptions of $E(\mathbf{e}_1)$, $E(\mathbf{e}_2)$ and $E(y)$ from the FHE oracle.\footnote{To be precise, asking the oracle for the encryption of an $n$-element vector actually means asking the oracle for $n$ encryptions, once per element. For succinctness, we omit this detail.} $\cal{B}$ gives $E(\mathbf{e}_1)$, $E(\mathbf{e}_2)$, $E(y)$ and $E(m_b)$ to $\cal{A}_\text{strong}$ as the description of the transformed network function $\phi$. Note that this is essentially the policy
\[
\texttt{if } x_1 == y \texttt{ then } x_2 \leftarrow m_b.
\]
During the test state, whenever $\cal{A}_\text{strong}$ asks for the result (encryptions from packet processing) of a packet $\mathbf{x}$ under PNFV, $\cal{B}$ does as follows. If $x_1 = y$, $\cal{B}$ asks the FHE oracle for the encryptions of $x_1, x_3, \ldots, x_n$. It further requests the oracle for the encryption of $0$, and upon receiving $E(0)$, asks for the encryption of $E(m_b) + E(0) = E(m_b)'$. $\cal{B}$ constructs the vector
\[
E(\mathbf{x}') = 
\begin{pmatrix}
E(x_1) & E(m_b)' & E(x_3) & \cdots & E(x_n)
\end{pmatrix}.
\]
Otherwise if $x_1 \ne y$, $\cal{B}$ asks for the encryption of $x_2$ from the FHE oracle and $E(m_b)'$ is replaced with $E(x_2)$ in the above vector. This vector is then given to $\cal{A}_\text{strong}$. When $\cal{A}_\text{strong}$ submits its guess for $\psi$ as the tuple $(y',z')$ (as a match-action pair), $\cal{B}$ does as follows. If $y' = y$ and $z' = z_0$, then $\cal{B}$ outputs $b = 0$ as its guess, i.e., $\cal{B}$ guesses that $E(m_b)$ is the encryption of $m_0 = z_0$. Otherwise, if $y' = y$ and $z' = z_1$, then $\cal{B}$ outputs $b = 1$, i.e., $\cal{B}$ guesses that $E(m_b)$ is the encryption of $m_1 = z_1$. To see that this strategy works, notice that if $m_b = m_0 = z_0$, the above policy is
\[
\texttt{if } x_1 == y \texttt{ then } x_2 \leftarrow z_0,
\]
and if $m_b = m_1 = z_1$, the above policy is
\[
\texttt{if } x_1 == y \texttt{ then } x_2 \leftarrow z_1,
\]
as required.
\end{argument}

\begin{claim}
If the BGN cryptosystem $(E, D)$ is semantically secure (indistinguishable under chosen plaintext attack), then the PNFV scheme based on the BGN scheme is private against $\cal{A}_\text{strong}$.
\end{claim}

\begin{argument}
The proof is similar to above with minor differences, which we highlight here. For network function transformation, $\cal{B}$ asks the BGN oracle the encryption of $n-1$ zeros and constructs the vector 
\[
E(m_b\mathbf{e}_2) = 
\begin{pmatrix}
E(0)' & E(m_b) & E(0)'' & \cdots & E(0)^{(n-1)}
\end{pmatrix}.
\]
It then asks the oracle for encryptions of $E(1)$, $E(\mathbf{e}_1)$, $E(y)$ and $E(\mathbf{e}_2)$, and sends the tuple
\[
(E(1), E(\mathbf{e}_1), E(y), E(\mathbf{e}_2), E(m_b\mathbf{e}_2))
\] 
to $\cal{A}_\text{strong}$ as the description of the transformed network function. During the guess state of $\cal{A}_\text{strong}$, whenever a packet $\mathbf{x}$ is presented to $\cal{B}$, it asks the BGN oracle for the encryption of $1 - x_1 + y$ and labels the resulting encryption as $E(m(\mathbf{x}))$ (note that if $x_1 = y$ then the matching function is simply the encryption of $1$). For the action function, $\cal{B}$ asks the FHE oracle for the encryptions of $x_1, x_3, \ldots, x_n$. It further requests the oracle for the encryption of $0$, and upon receiving $E(0)$, asks for the encryption of $E(m_b) + E(0) = E(m_b)'$. $\cal{B}$ constructs the vector
\[
E(a(\mathbf{x})) = 
\begin{pmatrix}
E(x_1) & E(m_b)' & E(x_3) & \cdots & E(x_n)
\end{pmatrix}.
\]
Finally $\cal{B}$ asks the BGN oracle for the encryption of $E(\mathbf{x})$. It sends $E(\mathbf{x})$, $E(m(\mathbf{x}))$ and $E(a(\mathbf{x}))$ to $\cal{A}_\text{strong}$.
\end{argument}

\begin{claim}
If the probabilistic encryption scheme $E$ is semantically secure, the PEKS scheme is semantically secure against the chosen keyword (plaintext) attack, its trapdoor function $T$ is not invertible, and the pseudorandom permutation $\sigma$ is indistinguishable from a random permutation, then the PNFV scheme described in Section~\ref{sub:weak-peks} is private against $\cal{A}_\text{weak}$.
\end{claim}

\begin{argument}
We define the following statements:
\begin{itemize}
	\item $D$: the PNFV scheme described in Section~\ref{sub:weak-peks} is private against $\cal{A}_\text{weak}$.
	\item $A_1$: the probabilistic encryption scheme $E$ is semantically secure.
	\item $A_2$: the PEKS scheme is semantically secure against the chosen keyword (plaintext) attack and the trapdoor function $T$ is not invertible.
	\item $A_3$: the pseudorandom permutation $\sigma$ is indistinguishable from a random permutation.
\end{itemize}
We further refine $D$ as follows:
\begin{itemize}
	\item $D_1$: $\cal{A}_\text{weak}$ does not know the tuple $(j, z)$.
	\item $D_2$: $\cal{A}_\text{weak}$ does not know the tuple $(i, j, y)$.
	\item $D_3$: $\cal{A}_\text{weak}$ does not know the tuple $(i, j)$.
\end{itemize}
Then it follows that:
\[
D \Leftrightarrow D_1 \wedge D_2 \wedge D_3.
\]
That is, the PNFV scheme is not private if $\cal{A}_\text{weak}$ knows any of the aforementioned tuples. The claim states that
\[
 A_1 \wedge A_2 \wedge A_3 \Rightarrow D
\]
or equivalently 
\begin{equation}
\label{eq:log-neg}
\neg D \Rightarrow \neg A_1 \vee \neg A_2 \vee \neg  A_3 .
\end{equation}
In the following, in a series of ``games'' we show that for $i, j, k \in \{1, 2, 3\}$ and $i \ne j \ne k$,
\[
\neg D_i \wedge A_j \wedge A_k \Rightarrow \neg A_i.
\]
The conjunction of the above propositions is equivalent to proposition~\ref{eq:log-neg}, since
\begin{align*}
&\bigwedge_i \left( \neg D_i \wedge A_j \wedge A_k \Rightarrow \neg A_i \right) \\
&\Leftrightarrow \bigwedge_i \left( D_i \vee \neg A_j \vee \neg A_k \vee \neg A_i \right) \\
																									&\Leftrightarrow \bigwedge_i \left( D_i \vee \neg A_1 \vee \neg A_2 \vee \neg A_3 \right) \\
																								   &\Leftrightarrow \left(D_1 \wedge D_2 \wedge D_3 \right) \vee \left(\neg A_1 \vee \neg A_2 \vee \neg A_3 \right) \\
																								   &\Leftrightarrow D \vee \left(\neg A_1 \vee \neg A_2 \vee \neg A_3 \right) \\
																								   & \Leftrightarrow \neg D \Rightarrow \neg A_1 \vee \neg A_2 \vee \neg  A_3,
\end{align*}
where we have implicitly used the tautology $P \Rightarrow Q \Leftrightarrow \neg P \vee Q$. For notational convenience, we shall use $\sigma(\mathbf{x})$ to denote the permuted vector after the application of the permutation $\sigma$. On the other hand, $\sigma(x)$ shall denote the permutation of single element $x \in \mathbf{x}$ under $\sigma$. We shall denote by $I$ the vector of indexes $\{1, 2, \ldots, n\}$. The notation $\mathbf{x} || I $ denotes the vector whose $l$th element is $x_l || l $. 

\medskip
\noindent\textit{Game 1.} Suppose $A_2$ and $A_3$ hold. Then if $\cal{A}_\text{weak}$ learns the tuple $(j, z)$ in the PNFV scheme, then the probabilistic encryption scheme $E$ is not semantically secure, i.e., $\neg D_1 \Rightarrow \neg A_1$. 

We construct an adversary $\cal B$ that uses $\cal{A}_\text{weak}$ as a subroutine. $\cal B$ issues the challenger with $z_0 || 2$ and $z_1 || 2$ as the two plaintexts it wants to be challenged on. Let $m_0 = z_0 || 2$ and $m_1 = z_1 || 2$. The challenger returns $E(m_b)$ to $\cal{B}$ such that $b = 0$ with probability $\frac{1}{2}$. $\cal B$ samples two uniform random bit strings with length equal to the range of the trapdoor function $T$, and labels these values $T(y || 1)$ and $T(2)$. Note that these are not actual trapdoors, but random values (dummy trapdoors) simulating the behaviour of a non-invertible trapdoor. $\cal {B}$ gives $E(m_b)$, $T(y || 1)$ and $T(2)$ to $\cal{A}_\text{weak}$. Whenever $\cal{A}_\text{weak}$ asks for new packet encryptions, $\cal B$ samples a packet $\mathbf{x}$ from the public distribution $\cal{D}$. If $x_1 = y$ for a predetermined and fixed value of $y$, $\cal B$ asks the $E$ oracle for $n-1$ encryptions of $x_l || l$ such that $l \ne 1$, and an encryption of $0$ followed by the encryption of $E(m_b)' = E(0) + E(m_b)$, and constructs $E(\mathbf{x} || I)$, such that $E(x_1 || 1) = E(m_b)'$. $\cal{B}$ then randomly generates a permutation $\sigma$ and permutes $E(\mathbf{x} || I)$ obtaining $\sigma(E(x || I))$. Note that this permutation $\sigma$ is generated by $\cal{B}$ itself. $\cal{B}$ also generates $2n$ random bit strings of size equal to the range of $\mathcal{E}$. $n$ of these values are used to simulate $\sigma(\mathcal{E}(\mathbf{x} || I))$, and the other $n$ to simulate $\sigma(\mathcal{E}(I))$. $\cal{B}$ gives these permuted encryptions to $\cal{A}_\text{weak}$. To simulate the $\mathsf{test}$ routine, if $x_1 = y$, $\cal B$ gives $\sigma(1)$ and $\sigma(2)$ to $\cal{A}_\text{weak}$, i.e., the permuted indexes corresponding to the match and action. $\cal B$ further replaces $\sigma(E(x_2 || 2))$ with $\sigma(E(m_b))$ in $\sigma(E(\mathbf{x} || I))$, and gives the resultant $\sigma(E(\mathbf{x} || I))$ to $\cal{A}_\text{weak}$. Otherwise it simply gives $\sigma(E(\mathbf{x} || I))$ to $\cal{A}_\text{weak}$ (without replacing $\sigma(E(x_2 || 2))$). When $\cal{A}_\text{weak}$ outputs $(j', z')$ as its guess for the policy, $\cal{B}$ outputs $0$ if $z' = z_0$; otherwise if $z'=z_1$, $\cal{B}$ outputs $1$. 

\medskip
\noindent\textit{Game 2.} Suppose $A_1$ and $A_3$ hold. Then if $\cal{A}_\text{weak}$ learns the tuple $(i, j, y)$ in the PNFV scheme, then the PEKS scheme is not semantically secure against the chosen keyword (plaintext) attack or the trapdoor function $T$ is invertible, i.e., $\neg D_2 \Rightarrow \neg A_2$. 

We show this in two sub-games. 

\medskip
\noindent\textit{Game 2.1.} Suppose $T$ is not invertible, then if $\cal{A}_\text{weak}$ learns the tuple $(y, i, j)$ in the PNFV scheme, the PEKS scheme is not semantically secure against the chosen keyword (plaintext) attack.

We consider an adversary $\cal B$ who chooses $m_0$ and $m_1$ as two chosen keywords (plaintexts) and is given $\mathcal{E}(m_b)$ such that $b = 0$ with probability $\frac{1}{2}$. $\cal B$ has to guess $b$. It can ask the challenger for further encryptions of any plaintext. $\cal B$ is also given access to two instances of $\mathsf{test}$ oracle; one, labelled $\mathsf{test}_0$, instantiated with the trapdoor $T(m_b)$ and the other, labelled $\mathsf{test}_1$, with the trapdoor $T(j)$, where $j$ is chosen by $\cal B$. Note that $\cal B$ is not given the trapdoor values themselves. We assume an oracle $\cal P$ which when invoked, generates a random $\mathbf{x}$ according to the distribution $\cal{D}$, and outputs $\sigma({E}(\mathbf{x} || I))$, $\sigma(\mathcal{E}(\mathbf{x} || I))$ and $\sigma(\mathcal{E}({I}))$, where $\sigma({E}(\mathbf{x} || I))$ is a vector of $n$ random bit strings each of length equal to the range of $E$ and $\sigma$ is a (truly) random permutation. More specifically, $\cal P$ is also given oracle access to $\mathcal{E}$. Our adversary $\cal B$ again uses $\cal{A}_\text{weak}$ for the rescue. It chooses $m_0 = y_0 ||1$ and $m_1 = y_1 || 1$ as its two chosen plaintexts. Upon receiving $\mathcal{E}(m_b)$, it generates two random bit strings of length equal to the range of $T$. One of these simulates $T(m_b)$ and the other $T(j) = T(2)$. $\cal B$ initializes the $\mathsf{test}_1$ oracle with $j = 2$. $\cal B$ also samples a bit string uniformly at random to simulate $E(z||j)$ (with length equal to the range of $E$). It gives these simulations of ${T}(m_b)$, $T(3)$ and $E(z||j)$ to $\cal{A}_\text{weak}$. In the testing phase, $\cal B$ queries the $\cal P$ oracle and obtains $\sigma({E}(\mathbf{x}) || I)$, $\sigma(\mathcal{E}(\mathbf{x} || I))$ and $\sigma(\mathcal{E}({I}))$ as a result, and duly sends them to $\cal{A}_\text{weak}$. It also runs the $\mathsf{test}_0$ oracle to determine if there is a match, and if yes replaces the value in $\sigma({E}(\mathbf{x} || I))$ corresponding to the output of the oracle $\mathsf{test}_1$ with $E(z||j)$. When $\cal{A}_\text{weak}$ outputs $(i', j', y')$, $\cal B$ outputs $b = 0$ if $y' = y_0$. Else if $y' = y_1$, $\cal B$ outputs $b = 1$.  

\medskip
\noindent\textit{Game 2.2.} Suppose the PEKS scheme is semantically secure against the chosen keyword (plaintext) attack, then  if $\cal{A}_\text{weak}$ learns the tuple $(i, j, y)$ in the PNFV scheme, the trapdoor function $T$ is invertible.

This is similar to above. This time, instead of $\mathcal{E}(m_b)$, $\cal B$ is given $T(m_b)$. Note that if $T$ is invertible, then finding $b$ is straightforward. $\cal B$ chooses  $m_0 = y_0 ||1$ and $m_1 = y_1 || 1$ as before, and further asks for the trapdoor of $2$ and gets $T(2)$ as a result (where $j = 2$ is the instantiation of $j$). $\cal B$ can ask for any further trapdoors pertaining to the condition that the keyword should not equal $m_0$ or $m_1$. We now also have an oracle $\mathcal{E}$ which upon asked for the encryption of some plaintext $x$ returns a uniform random value in the range of $\mathcal{E}$. The oracle keeps the record of the value of $\mathcal{E}(x)$ against $x$ in a table. This oracle can also be accessed by the $\cal P$ oracle and the $\mathsf{test}$ oracle (we have only one $\mathsf{test}$ oracle this time). At the end, $\cal B$ checks the output of $\cal{A}_\text{weak}$ obtained as $(i', j', y')$, and returns the bit $b$ as before.

\medskip
\noindent\textit{Game 3.} Suppose $A_1$ and $A_2$ hold. Then if $\cal{A}_\text{weak}$ learns the tuple $(i, j)$ in the PNFV scheme, then the pseudorandom permutation $\sigma$ is distinguishable from a random permutation, i.e., $\neg D_3 \Rightarrow \neg A_3$. 

We assume the following challenge game between $\cal B$ and $\sigma$. $\cal B$ can invoke $\sigma$ as many times as it wants by making a call with the query `\texttt{next}'. Each such call will be called an iteration of $\sigma$. Note that before the first call, it is presumed that $\sigma$ is in the identity configuration, i.e., $(1, 2, \ldots, n)$. $\cal B$ can choose an integer $u \in \{1, 2, \ldots, n\}$ and give it to the challenger. The challenger chooses another integer $m \in \{1, 2, \ldots, n\}$ such that $m \ne u$, which $\cal B$ has to guess. For each oracle call to $\sigma$, $\cal B$ can ask for the permutation of the fixed integer $u$ as well as $\sigma(m)$ (i.e., the permuted value of the unknown integer $m$). $\cal B$ has to determine $m$. Note that if $\sigma$ is indistinguishable from a random permutation then the guess of $\cal B$ should be no better than $\frac{1}{n - 1}$. Suppose $\cal B$ chooses $u = 1$. $\cal B$ gives random values to the adversary $\cal{A}_\text{weak}$ to substitute ${T}(y || i)$, $T(j)$, $E(z||j)$, $\sigma({E}(\mathbf{x} || I))$, $\sigma(\mathcal{E}(\mathbf{x} || I))$ and $\sigma(\mathcal{E}({I}))$, where the packet $\mathbf{x}$ is generated by $\cal B$ according to the public distribution $\cal D$. Whenever $x_1 = y$, $\cal B$ invokes $\sigma$, and asks for $\sigma(1)$ and $\sigma(m)$. $\cal B$ then replaces $\sigma(E({x}_m || m))$ with $E(z||j)$. Whenever $\cal{A}_\text{weak}$ outputs $(i', j')$, $\cal B$ outputs $m = j'$.
\end{argument}


\begin{thebibliography}{10}

\bibitem{relic-toolkit}
D.~F. Aranha and C.~P.~L. Gouv\^{e}a.
\newblock {RELIC is an Efficient LIbrary for Cryptography}.
\newblock \url{https://github.com/relic-toolkit/relic}.

\bibitem{argyraki2010verifiable}
K.~Argyraki, P.~Maniatis, and A.~Singla.
\newblock {Verifiable network-performance measurements}.
\newblock In {\em ACM CoNEXT Conference}, 2010.

\bibitem{bloom1970space}
B.~H. Bloom.
\newblock Space/time trade-offs in hash coding with allowable errors.
\newblock {\em Communications of the ACM}, 13(7), 1970.

\bibitem{peks}
D.~Boneh, G.~Di~Crescenzo, R.~Ostrovsky, and G.~Persiano.
\newblock {Public key encryption with keyword search}.
\newblock In {\em Eurocrypt}, 2004.

\bibitem{ibe}
D.~Boneh and M.~K. Franklin.
\newblock {Identity-Based Encryption from the Weil Pairing}.
\newblock In {\em CRYPTO}, 2001.

\bibitem{bgn}
D.~Boneh, E.-J. Goh, and K.~Nissim.
\newblock {Evaluating 2-DNF Formulas on Ciphertexts}.
\newblock In {\em TCC}, 2005.

\bibitem{mlm-cryptanalysis}
J.~H. Cheon, K.~Han, C.~Lee, H.~Ryu, and D.~Stehle.
\newblock {Cryptanalysis of the Multilinear Map over the Integers}.
\newblock In {\em Eurocrypt}, 2015.

\bibitem{chow2009controlling}
R.~Chow, P.~Golle, M.~Jakobsson, E.~Shi, J.~Staddon, R.~Masuoka, and J.~Molina.
\newblock {Controlling data in the cloud: outsourcing computation without
  outsourcing control}.
\newblock In {\em ACM workshop on Cloud computing security}, 2009.

\bibitem{mlm-integers}
J.-S. Coron, T.~Lepoint, and M.~Tibouchi.
\newblock {Practical Multilinear Maps over the Integers}.
\newblock In {\em CRYPTO}, 2013.

\bibitem{freeman2010converting}
D.~M. Freeman.
\newblock {Converting pairing-based cryptosystems from composite-order groups
  to prime-order groups}.
\newblock In {\em EUROCRYPT 2010}. 2010.

\bibitem{waypoints}
G.~Gibb, H.~Zeng, and N.~McKeown.
\newblock {Outsourcing Network Functionality}.
\newblock In {\em HotSDN}, 2012.

\bibitem{fdd}
M.~G. Gouda and A.~X. Liu.
\newblock {Structured Firewall Design}.
\newblock {\em Comput. Netw.}, 2007.

\bibitem{haeberlen2010accountable}
A.~Haeberlen, P.~Aditya, R.~Rodrigues, and P.~Druschel.
\newblock {Accountable Virtual Machines.}
\newblock In {\em OSDI}, 2010.

\bibitem{nohype}
E.~Keller, J.~Szefer, J.~Rexford, and R.~B. Lee.
\newblock {NoHype: Virtualized Cloud Infrastructure Without the
  Virtualization}.
\newblock In {\em ISCA}, 2010.

\bibitem{bf-firewall}
A.~R. Khakpour and A.~X. Liu.
\newblock {First Step Toward Cloud-Based Firewalling}.
\newblock In {\em SRDS}, 2012.

\bibitem{sel4}
G.~Klein, K.~Elphinstone, G.~Heiser, J.~Andronick, D.~Cock, P.~Derrin,
  D.~Elkaduwe, K.~Engelhardt, R.~Kolanski, M.~Norrish, T.~Sewell, H.~Tuch, and
  S.~Winwood.
\newblock {seL4: Formal Verification of an OS Kernel}.
\newblock In {\em SOSP}, 2009.

\bibitem{luby-rackoff}
M.~Luby and C.~Rackoff.
\newblock {How to Construct Pseudorandom Permutations from Pseudorandom
  Functions}.
\newblock {\em SIAM J. Comput.}, 1988.

\bibitem{click-os}
J.~Martins, M.~Ahmed, C.~Raiciu, V.~Olteanu, M.~Honda, R.~Bifulco, and
  F.~Huici.
\newblock {ClickOS and the Art of Network Function Virtualization}.
\newblock In {\em NSDI}, 2014.

\bibitem{openflow}
N.~McKeown, T.~Anderson, H.~Balakrishnan, G.~Parulkar, L.~Peterson, J.~Rexford,
  S.~Shenker, and J.~Turner.
\newblock {OpenFlow: Enabling Innovation in Campus Networks}.
\newblock {\em SIGCOMM Comput. Commun. Rev.}, 2008.

\bibitem{handbook-ac}
A.~J. Menezes, S.~A. Vanstone, and P.~C.~V. Oorschot.
\newblock {\em {Handbook of Applied Cryptography}}.
\newblock 1st edition, 1996.

\bibitem{DBLP:conf/ccs/NaehrigLV11}
M.~Naehrig, K.~Lauter, and V.~Vaikuntanathan.
\newblock {Can homomorphic encryption be practical?}
\newblock In {\em CCSW}, 2011.

\bibitem{hey-you}
T.~Ristenpart, E.~Tromer, H.~Shacham, and S.~Savage.
\newblock {Hey, You, Get off of My Cloud: Exploring Information Leakage in
  Third-party Compute Clouds}.
\newblock In {\em ACM CCS}, 2009.

\bibitem{top26}
K.~Searl.
\newblock {Top 26 Companies in the Global NFV Market}.
\newblock
  \url{http://www.technavio.com/blog/top-26-companies-in-the-global-nfv-market},
  2014.

\bibitem{aplomb}
J.~Sherry, S.~Hasan, C.~Scott, A.~Krishnamurthy, S.~Ratnasamy, and V.~Sekar.
\newblock {Making Middleboxes Someone Else's Problem: Network Processing as a
  Cloud Service}.
\newblock In {\em SIGCOMM}, 2012.

\bibitem{blindbox}
J.~Sherry, C.~Lan, R.~A. Popa, and S.~Ratnasamy.
\newblock {BlindBox: Deep Packet Inspection over Encrypted Traffic}.
\newblock In {\em SIGCOMM}, 2015.

\bibitem{mlm-firewall}
J.~Shi, Y.~Zhang, and S.~Zhong.
\newblock {Privacy-preserving Network Functionality Outsourcing}.
\newblock \url{http://arxiv.org/abs/1502.00389}, 2015.

\bibitem{vu2013hybrid}
V.~Vu, S.~Setty, A.~J. Blumberg, and M.~Walfish.
\newblock {A hybrid architecture for interactive verifiable computation}.
\newblock In {\em IEEE S\&P)}, 2013.

\bibitem{hypersafe}
Z.~Wang and X.~Jiang.
\newblock {HyperSafe: A Lightweight Approach to Provide Lifetime Hypervisor
  Control-Flow Integrity}.
\newblock In {\em IEEE S\&P}, 2010.

\bibitem{cloud-visor}
F.~Zhang, J.~Chen, H.~Chen, and B.~Zang.
\newblock {CloudVisor: Retrofitting Protection of Virtual Machines in
  Multi-tenant Cloud with Nested Virtualization}.
\newblock In {\em SOSP}, 2011.

\bibitem{zhang2008packet}
X.~Zhang, A.~Jain, and A.~Perrig.
\newblock {Packet-dropping adversary identification for data plane security}.
\newblock In {\em ACM CoNEXT Conference}, 2008.

\bibitem{zhang-juels}
Y.~Zhang, A.~Juels, M.~K. Reiter, and T.~Ristenpart.
\newblock {Cross-VM Side Channels and Their Use to Extract Private Keys}.
\newblock In {\em ACM CCS}, 2012.

\end{thebibliography}
\end{document}